%% file: main.tex
\documentclass[10pt, journal]{IEEEtran}
\usepackage{amsmath,amssymb,amsfonts,stmaryrd}
\usepackage{makecell}
\usepackage{booktabs,multicol,multirow}
\usepackage{algorithm}
\usepackage{mathtools}
\usepackage{url}
\usepackage{soul}
\usepackage{cite}
\usepackage[dvipsnames]{xcolor}
\usepackage[noend]{algpseudocode}
\usepackage[hidelinks]{hyperref}
\usepackage{dblfloatfix}  
\usepackage{soul}

\usepackage{fontawesome}
\usepackage[inline]{enumitem}
\usepackage{threeparttable}
\usepackage[super]{nth}

\newcommand{\oursTwo}[1]{{\color{RoyalBlue}\textit{#1}}} 
\newcommand{\oursFive}[1]{{\color{ForestGreen}\textit{#1}}} 

\usepackage[  round-mode = places, round-precision = 2]{siunitx}

\newcommand{\blue}[1]{{\color{black}#1}}

\definecolor{green}{RGB}{0,128,0}
\definecolor{red}{RGB}{255,0,0}

\makeatletter
\newcommand*\titleheader[1]{\gdef\@titleheader{#1}}
\AtBeginDocument{%
  \let\st@red@title\@title
  \def\@title{%
    \bgroup\normalfont\normalsize\centering\@titleheader\par\egroup
    \vskip1ex\st@red@title}
}
\makeatother

\title{
Arbitrary Precision Printed Ternary Neural Networks with Holistic Evolutionary Approximation
}

\author{
Vojtech Mrazek,
Konstantinos Balaskas,
Paula Carolina Lozano Duarte,
Zdenek Vasicek,\\
Mehdi B. Tahoori,
Georgios Zervakis

\thanks{Manuscript received March 6, 2025, \blue{revised July 17, 2025}, accepted August 20, 2025. (\textit{Corresponding author: V. Mrazek, e-mail: mrazek@fit.vutbr.cz}).}%
\thanks{This work was supported by the Czech Science Foundation project
25-15490S, by the European Research Council (ERC, Grant No. 101052764), and by the H.F.R.I call “Basic research Financing (Horizontal support of all Sciences)” under the National Recovery and Resilience Plan “Greece 2.0” (H.F.R.I. Project Number: 17048).
}
\thanks{V. Mrazek and Z. Vasicek are with the Faculty of Information Technology, Brno University of Technology, 612 00 Brno, Czech Republic}
\thanks{P. C. Lozano Duarte and M.~B.~Tahoori are with the Department of Computer Science, Karlsruhe Institute of Technology, Karlsruhe 76131, Germany}
\thanks{K. Balaskas and G. Zervakis are with the Computer Engineering \& Informatics Dept., University of Patras, Patras 26504, Greece.}
}
\titleheader{\vspace{-25pt}\small{Accepted at IEEE Transactions on Circuits and Systems for Artificial Intelligence (TCASAI). DOI: 10.1109/TCASAI.2025.3604384. \textsuperscript{\textcopyright}IEEE}}

\begin{document}
\bstctlcite{IEEEexample:BSTcontrol}

\maketitle
\begin{abstract}
Printed electronics offer a promising alternative for applications beyond silicon-based systems, requiring properties like flexibility, stretchability, conformality, and ultra-low fabrication costs.
Despite the large feature sizes in printed electronics, printed neural networks have attracted attention for meeting target application requirements, though realizing complex circuits remains challenging.
This work bridges the gap between classification accuracy and area efficiency in printed neural networks, covering the entire processing-near-sensor system design and co-optimization from the analog-to-digital interface--a major area and power bottleneck--to the digital classifier.
We propose an automated framework for designing printed Ternary Neural Networks with arbitrary input precision, utilizing multi-objective optimization and holistic approximation.
Our circuits outperform existing approximate printed neural networks by 17x in area and 59x in power on average, being the first to enable printed-battery-powered operation with under 5\% accuracy loss while accounting for analog-to-digital interfacing costs.
\end{abstract}


\input{introduction}

\input{background}

\input{tnn}

\input{approx}

\input{experimental}
\input{conclusion}


\input{bios}

\vfill

\end{document}

%% file: introduction.tex
\section{Introduction}\label{sec:introduction}

Over the past several decades, the \blue{Very-Large-Scale Integration (VLSI) circuit} industry has managed to fuel Moore's law, and feature scaling is expected to persist for the next ten years and beyond~\cite{Gargini:MED2023}.
The common theme among these technological innovations is the never-ending pursuit of improving the power-performance-area (PPA) of transistors, and silicon-based systems overall.
However, numerous application domains fall outside the PPA target, including forensics~\cite{salivary:Talanta:2020}, disposables and smart packaging~\cite{smartpackaging2022, disposable:JSNB:2023}, accessible healthcare products and wearables~\cite{bodytemperature:sna:2020,pressuresensor:research:2022,wearable:adma:2022,healthcare:Nanoscale:2024}, and the ten trillion-dollar fast-moving consumer goods market~\cite{Bleier:ISCA:2020:printedmicro}.
Such applications require \textit{ultra-low-cost fabrication, along with stretchability, porosity, flexibility, and conformality}---demands that rigid silicon systems cannot meet, positioning printed electronics at the forefront for integrating smart services in these domains that have yet to witness significant computing infiltration.

Printed electronics use mask-less, additive methods like jet, screen, or gravure printing, achieving ultra-low cost (even sub-cent) and fast-turnaround fabrication~\cite{cui2016printed, chang2017circuits}
\blue{(see Section~\ref{subsec:pe} for more information).
Coupled with low capital equipment costs, printed electronics enable rapid and inexpensive manufacturing.}\label{rev:r3c9}
On the other hand, the low-resolution printing results in large feature sizes, low integration density, and high device latencies~\cite{lei2019low, cadilha2017digital}.
These limitations pose challenges for implementing printed Machine Learning (ML) classifiers \blue{(i.e., complex datapaths with increased gate count)}, but are essential for numerous applications in the aforementioned domains~\cite{Mubarik:MICRO:2020:printedml}.
These applications frequently involve classifying analog sensor data to derive meaningful insights.

Extensive research has been dedicated to printed neural networks~\cite{Mubarik:MICRO:2020:printedml,Weller:2021:printed_stoch,Armeniakos:DATE2022:axml,Armeniakos:TC2023:codesign,Afentaki:ICCAD2023:axmac,Afentaki:DATE2024:gatrain,mrazek:iccad:2024}.
To address the inherent limitations of printed electronics---particularly in terms of area and power---two main design paradigms gain interest in printed ML: bespoke~\cite{Kumar2017Bespoke} and approximate~\cite{Jiang:JPROC2020}.  
Bespoke design involves fully customized circuits tailored to each ML model and dataset.
This is enabled by the ultra-low fabrication and non-recurring engineering (NRE) costs of printed electronics, achieved by hardwiring model parameters into the circuit.
Such a level of customization is unattainable in silicon-based systems.  

Efficiency can be further enhanced by leveraging the error resilience of ML applications and employing approximate computing~\cite{Henkel:ICCAD2022:expedition}.  
\textit{Printed electronics and ML synergistically act as enablers for approximate computing, and vice versa.}  
Printed electronics are subject to strict physical constraints, such as tight area limits and restricted power availability, and can integrate only a very limited number of devices.  
\blue{The problem is exacerbated when ML circuits are considered, which inherently have significant hardware overheads.
Approximation techniques that aggressively reduce the gate count are favored to enable practical implementations and deliver ultra-area-efficient solutions.}\label{rev:r3c10}
On the other hand, extensive research in approximate computing emerged about fifteen years ago, but it saw limited adoption in real-world applications.
This is largely due to the poor generalization of approximate circuits and the high fabrication costs of silicon systems, which discourage designers from creating systems prone to often significant or unforeseen errors or incurring the additional costs of verifying approximate designs.  
In contrast, as aforementioned, printed electronics offer negligible fabrication costs, fast on-field manufacturing, and a typically short circuit lifespan---often just up to a few days.  
These advantages empower designers to experiment with and adopt unconventional computing paradigms, such as approximate computing.  
\blue{This is because, unlike traditional silicon systems where errors can lead to costly failures and re-fabrication, the low-risk context of printed circuits makes approximation-induced inaccuracies far more acceptable.
}\label{rev:r3c13}
Additionally, bespoke design in printed electronics enables ML circuit-specific approximations, addressing the generalization challenge of approximate circuits and maximizing hardware savings.  
Finally, the relatively low complexity of target printed ML applications makes the corresponding circuits even more resilient to errors~\cite{Armeniakos:CSUR2022}, enabling more aggressive approximations and enhancing the applicability and benefits of approximate computing. 

In this work, we focus on designing digital approximate printed neural networks using Electrolyte-Gated FET (EGFET) technology~\cite{Bleier:ISCA:2020:printedmicro}, which offers good mobility characteristics and enables operation at low supply voltages~\cite{Marques:Materials:2019}, making it ideal for printed battery-powered applications.
Recognizing the prohibitive cost of printed multipliers~\cite{Armeniakos:DATE2022:axml,Armeniakos:TC2023:codesign}, the state of the art has shifted to multiplier-less networks, such as power-of-2 quantized Multilayer Perceptrons (MLPs)~\cite{Afentaki:ICCAD2023:axmac,Afentaki:DATE2024:gatrain}
or Ternary Neural Networks (TNNs)~\cite{mrazek:iccad:2024}.
This allows bespoke circuit implementations to use only adders/subtractors and re-wiring for shift operations.
However,~\cite{Armeniakos:DATE2022:axml,Armeniakos:TC2023:codesign,Afentaki:ICCAD2023:axmac,Afentaki:DATE2024:gatrain} require higher precision inputs to maintain acceptable accuracy, \textit{overlooking the often dominant overheads of necessary analog-to-digital converters (ADCs) to process sensor data~\cite{Afentaki:ESL2024:adc, Duarte:ASPDAC2025}}.
Using $1$-bit inputs and simpler analog-to-binary converters (ABCs), in~\cite{mrazek:iccad:2024}, we avoided interfacing costs.
Existing approximate printed neural networks offer non-favorable accuracy-area trade-offs.
The MLPs in~\cite{Armeniakos:DATE2022:axml,Armeniakos:TC2023:codesign} offer low accuracy loss (e.g., $1$-$2$\%) but yield mainly modest hardware savings, with poor scalability for larger losses, while the multiplier-less $1$-bit TNNs~\cite{mrazek:iccad:2024} and power-of-2 MLPs~\cite{Afentaki:ICCAD2023:axmac,Afentaki:DATE2024:gatrain} often suffer from high accuracy loss (e.g., $>5$\%).

This paper extends our prior work in~\cite{mrazek:iccad:2024} and addresses the limitations mentioned above by \textit{co-optimizing} the analog frontend (i.e., ADC) and digital classifier logic, proposing an automated framework for \textit{designing approximate TNNs with arbitrary input precision}.
Our approach refines the exploration of the accuracy-area trade-off in printed neural networks.
To maximize area efficiency, we design, for the first time, approximate linear threshold gates (LTGs) to replace the TNN hidden neurons and use approximate popcount units in the output ones.
Our multi-objective optimization identifies the optimal approximation for each neuron.
Unlike the state of the art, our framework consistently delivers area-efficient solutions across all datasets and accuracy thresholds examined, achieving on average $17$x lower area for similar accuracy.

\noindent
\textbf{To summarize, our novel contributions in this work are}:
\begin{enumerate}[topsep=0pt,leftmargin=*]
    \item We are the first to investigate arbitrary input precision printed TNNs, co-designing and optimizing both the analog-to-digital interface and the digital classifier.
    \item We propose a systematic methodology for approximating LTGs and a comprehensive TNN approximation\footnote{\url{https://github.com/ehw-fit/arbitrary-input-tnn}.
   } using our multi-objective optimization framework that incorporates our precise area model alongside our approximate LTG and popcount units. Additionally, we leverage formal verification and Binary-Decision Diagrams to enable the design and error analysis of large approximate LTG circuits.
    \item For up to $5$\% accuracy loss, we enable, for the first time, printed-battery-powered operation of digital printed neural networks across all examined datasets, accounting also for the cost of analog-to-digital interfacing.
\end{enumerate}

%% file: background.tex
\blue{
\section{Background \& Related Work}
\label{sec:background_and_related}

\subsection{Printed Electronics}
\label{subsec:pe}

Printed electronics denote a class of fabrication techniques in which conductive materials are deposited onto flexible or rigid substrates using printing-based processes, such as inkjet, screen, and gravure printing~\cite{cui2016printed}.
Relying on resistor-nMOS logic, printed technology inherently differs from silicon-based CMOS by employing additive, maskless printing processes instead of complex, subtractive lithography-based fabrication, thus significantly reducing manufacturing costs.
Therefore, it enables the infiltration of computing into low-cost applications (e.g., healthcare wearables, fast-moving consumer goods) where attributes such as form factor, porosity, conformality and non-toxicity are prioritized.
This work employs the EGFET technology, which leverages electric double layer formation at the semiconductor–electrolyte interface to achieve high gate capacitance and sub-1V operation.
Using inkjet-printed indium oxide channels and printed electrolytes, EGFETs combine high carrier mobility with mechanical flexibility, making them ideal for low-power, battery-operated printed systems~\cite{Bleier:ISCA:2020:printedmicro}.

Building on these developments, printed neural networks--especially fully connected multilayer perceptron (MLP) circuits realized in printed electronics--have emerged as a way of embedding ML into ultra-low-cost systems.
These networks form the core of printed ML applications, which target resource-constrained domains such as smart packaging, diagnostic strips, and wearable healthcare devices.
In this context, terms like printed multipliers and digital approximate printed neural networks refer to specific circuit components or optimized implementations within these systems, all built using the same printed technology.
ML essentially enhances such printed systems with decision-making capabilities, increasingly critical in targeted far-edge applications.
}

\input{related}

%% file: related.tex

\blue{\subsection{Related Work}
\label{subsec:related}}
Printed electronics frequently targets applications centered on classification tasks.
This focus, combined with the expanding fast-moving consumer goods market, has spurred research into printed machine learning accelerators.
Various efforts have explored neural network models in this context, mostly favoring bespoke designs.


Bespoke circuits are customized for specific classifier models trained on particular datasets, eliminating generality-related overheads~\cite{Kumar2017Bespoke}.
This specialization extends to developing and implementing tailored approximation techniques~\cite{Jiang:JPROC2020}.
For instance, Armeniakos et al.~\cite{Armeniakos:DATE2022:axml} proposed post-training weight replacement with more hardware-friendly approximates to reduce the cost of the underlying multipliers.
In subsequent works, Armeniakos et al.~\cite{Armeniakos:TC2023:codesign} incorporated hardware-friendly weight approximation into the training process and implemented post-training addition approximation using simple truncation.
Afentaki et al.~\cite{Afentaki:ICCAD2023:axmac, Afentaki:DATE2024:gatrain} constrained weights to powers~of~2---leveraging that, in bespoke circuits, a multiplication by a power~of~2 is implemented simply by rewiring---approximated accumulations by pruning the adder trees, and employed bounded low-precision ReLu activation.
The authors in~\cite{Afentaki:ESL2024:adc,Duarte:ASPDAC2025} optimize the sensor-processor interface by reducing input representations and pruning the ADC circuit.
However, they do not explore classifier design optimizations, and high input precision is primarily needed to enable ADC pruning.
Lastly, an alternative approximation method, stochastic computing neural networks, was explored by Weller et al.~\cite{Weller:2021:printed_stoch} that frequently leads, however, to an unacceptable drop in accuracy.

The work presented here distinguishes itself from existing literature in several ways: it avoids the potential for significant accuracy losses associated with stochastic computing approaches; it employs a multiplier-free design and approximates LTGs and popcounts towards a holistic approximation; it is purely digital, avoiding the noise and variability issues that can affect low-resolution printed analog designs
~\cite{Zhao:DATE2023};
it considers arbitrary input precision and it begins optimization at the sensor boundary, rather than post-ADC.

%% file: tnn.tex
\section{Arbitrary Precision Printed TNNs}\label{sec:tnn}

This section details the design of our bespoke exact TNN implementation, which serves as the baseline for our approximation in Section~\ref{sec:axtnn}, and briefly discusses the cost of the printed ADCs needed to support arbitrary precision.

\begin{figure}[t!]
    \centering
    \includegraphics[width=0.8\columnwidth]{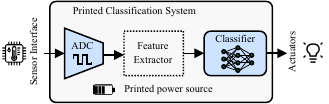}
    \caption{\blue{Overview of a printed classification-based system.}
    }
    \label{fig:pe_system}
\end{figure}

\blue{\figurename{}~\ref{fig:pe_system} presents an overview of a classification-based system in printed electronics, as would be deployed in a real-world printed ML application. 
In our work, we address the sensor interfacing and classifier blocks, by designing arbitrary-precision ADCs and approximate TNNs, respectively.}

\begin{figure}[t!]
\centering
\includegraphics[width=\columnwidth]{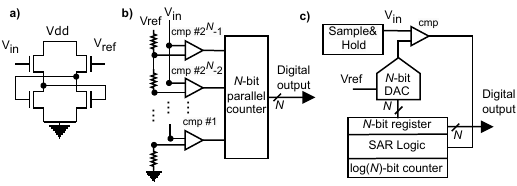}
\caption{a) Comparator in EGFET, b) $N$-bit Flash ADC and c) $N$-bit SAR ADC. EGFET is a resistor-nMOS only technology.}
\label{fig:Flash+SARADC}
\end{figure}

\input{table_adc}

\subsection{Printed ADCs}\label{subsec:adc}
Research on ADC design in printed electronics, particularly in EGFET technology, remains limited.
Therefore, we design and evaluate two popular ADC architectures: Flash~\cite{Kumary:FlashADC} and Successive Approximation Register (SAR)~\cite{SAR}.
We also evaluated Binary~\cite{Bhai:BinADC} and Sigma-Delta~\cite{Bajdechi:SigmaADC} ADCs; however, they are not functional in EGFET technology due to factors like high parasitic effects and the absence of pMOS transistors.

Flash ADCs (Fig.~\ref{fig:Flash+SARADC}b) are ideal for printed applications due to their regularity and simplicity. They operate by comparing the input voltage to multiple reference voltages using an array of analog comparators.
\blue{The comparator, i.e., the basic building block of the Flash ADC, is presented in Fig.~\ref{fig:Flash+SARADC}a.}\label{rev:r3c14}
This parallel, regular structure allows easy implementation.
The comparators' output represents the digital value in thermometer code, which is then converted to a binary encoding by counting the number of `1's with a digital parallel counter.
An $N$-bit ADC requires $2^N-1$ analog comparators.

SAR ADCs (Fig.~\ref{fig:Flash+SARADC}c) sequentially compare the input signal to reference levels, adjusting bit-by-bit, from most to least significant bits.
They require only one analog comparator, but also need a Digital-to-Analog Converter (DAC) and more complex digital logic (SAR logic) to store intermediate results and set reference values.
Moreover, SAR ADCs require registers that are highly area-inefficient in EGFET.
Hence, in SAR logic we use only an $N$-bit register for the output and a $\log_2N$ counter, with all other signals generated through counter multiplexing.

Table~\ref{tab:adc} reports the hardware overheads of the examined ADCs while the cost of an ABC~\cite{mrazek:iccad:2024} is $0.005$mm$^2$ and $0.001$mW.
Measurements are obtained using the EGFET \blue{Process Design Kit (PDK)}~\cite{Bleier:ISCA:2020:printedmicro} and Cadence Virtuoso.
For the examined precisions, Flash outperforms SAR in EGFET and will be used next.

\subsection{Arbitrary Precision Bespoke TNN Circuit}\label{subsec:tnn}

Focusing on area efficiency, a primary constraint in printed circuits~\cite{Mubarik:MICRO:2020:printedml}, we use a single hidden layer with input precision ranging from $1$ to $4$ bits.
\blue{Low-bit inputs allow for significantly smaller and less power-intensive ADCs, while also reducing the input precision requirements of the classifier (i.e., TNN).
Throughout recent research efforts in ML circuits for printed electronics~\cite{Armeniakos:DATE2022:axml,Armeniakos:TC2023:codesign,Duarte:ASPDAC2025,Armeniakos:DATE2024:dt}, low input precision (i.e., below 4 bits) has been commonly used since target domains have low performance and precision needs, enabling printed circuits with acceptable area and power overheads~\cite{Henkel:ICCAD2022:expedition}.
}\label{rev:r3c5}

The \texttt{sign} activation function is used for the hidden layer, and the output layer uses \texttt{argmax}.
The \texttt{sign} function computes the sign of the weighted sum of each hidden neuron, with \texttt{sign}($0$)$=$$1$~\cite{BinaryConnect,Binarized}, while the \texttt{argmax} function identifies the output neuron with the highest value.
Ternary weights $\{-1,0,1\}$ are used in both the hidden and output layers.

\begin{figure}[t!]
    \centering
    \includegraphics[width=0.8\columnwidth]{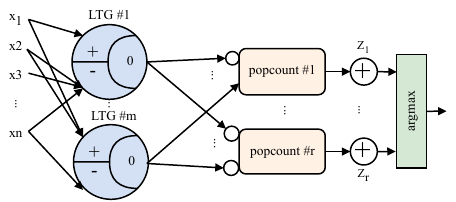}
    \caption{\blue{Bespoke exact TNN circuit overview.}}
    \label{fig:tnn}
\end{figure}

An abstract overview of our exact bespoke TNN circuits is illustrated in Fig.~\ref{fig:tnn}.
We adopt a fully parallel bespoke architecture, as it has proven more efficient than sequential and/or conventional ones for printed ML circuits~\cite{Mubarik:MICRO:2020:printedml}. 
\blue{Fully parallel refers to unfolded architectures, with all computations being performed simultaneously (i.e., purely combinational architecture).
Bespoke refers to highly customized implementations, specific to the model and dataset, where coefficients are hardwired in the circuit implementation.
Such a degree of customization would be infeasible in silicon-based conventional designs (i.e., non-bespoke), due to high fabrication and NRE costs.}\label{rev:r3c15}
In brief, each neuron has its own dedicated hardware, incorporating hardwired weights and eliminating the need for registers or memory that are highly inefficient in EGFET.
Connections with zero weights are removed from the corresponding neuron, while the sign of non-zero weights determines at design time whether the respective input is added or subtracted.

We implement the hidden neurons using LTGs and the output $y$ of each hidden neuron is given by:
\begin{equation}
    y = \begin{cases}
        1 & \text{if}\, \sum_{i=1}^{n} w_i x_i \geq 0 \\
        0 & \text{otherwise,}
    \end{cases}
\label{eq:ltg}
\end{equation}
where $w_i$ are the neuron weights, $x_i$ are the input activations, and $n$ is the number of inputs.
This reduces the precision required to represent $y$, with $-1$ being encoded as $0$ while $1$ remains $1$.
Note that~\eqref{eq:ltg} applies to any input precision, including $1$-bit inputs.

We implement the output neurons using popcount units. The output $o$ of each neuron is given by:
\begin{equation}   
o = \sum_{j=1}^{m} w_j y_j,
\label{eq:outputneuron}
\end{equation}
where $w_j$ are the neurons' weights and $y_j$ are the outputs of the hidden layer.
However, considering the aforementioned encoding for $y_j$, negation is achieved with a single NOT gate ($0\rightarrow1$ and $1\rightarrow0$).
In addition, due to the bespoke implementation, if a weight is $-1$, the corresponding activation is inverted before added to the respective popcount, whereas if the weight is $0$, the corresponding connection is removed.
The product of a zero weight by the respective activation is always $0$ (numerical value).
The (numerical) $0$ is in the middle of $-1$ and $1$ and, thus, in our encoding $0$ should be mapped to $\frac{1}{2}$ (i.e., $\frac{-1+1}{2}\rightarrow\frac{0+1}{2}$).
Therefore, a $\frac{1}{2}$ is missing from the result for each removed connection due to zero weight.
Hence, the output neuron can be implemented by a simple popcount unit (the sum in \eqref{eq:outputneuron} accumulates only zeros or ones) and to obtain the correct output, a constant correction term $\frac{Z}{2}$ must be added to the popcount result, where $Z$ equals the number of zero weights in that neuron.
Instead, we right-shift the popcount result $P$ and add $Z$: $2P + Z$.
Note that $Z$ is a constant known post-training.
Since the activation function of the output neuron is \texttt{argmax}, $2P + Z$ gives the same result as $P +\frac{Z}{2}$ without any rounding errors from integer division.
As a result, the output neuron computes the following:
\begin{equation}   
o = 2\!\sum_{\substack{j=1 \\ w_j\neq 0}}^{m}p_j+Z,\,\text{where}\,\, p_j=\begin{cases}
y_j,\, \text{if}\, w_j=1 \\
\overline{y_j},\, \text{if}\, w_j=-1 \\
\end{cases}.
\label{eq:outputneuron2}\vspace{-1ex}
\end{equation}
Similar to $w_j$, $Z$ values are also known at design time and are thus hardwired in the bespoke circuit implementation.

%% file: table_adc.tex
\begin{table}[t!]
\caption{Evaluation of ADCs in EGFET}\label{tab:adc}
\renewcommand{\arraystretch}{1.2}
\centering
\setlength{\tabcolsep}{4pt}
\begin{threeparttable}
\begin{tabular}{c|c|cc|c|cc|c|cc}

    \toprule
    \textbf{ADC} &
    \multirow{3}{*}{\rotatebox{90}{2-bit}} &
    \textbf{A}$^\dagger$ & \textbf{P}$^*$ & 
    \multirow{3}{*}{\rotatebox{90}{3-bit}} & 
    \textbf{A}$^\dagger$ & \textbf{P}$^*$ & 
    \multirow{3}{*}{\rotatebox{90}{4-bit}} & 
    \textbf{A}$^\dagger$ & \textbf{P}$^*$ 
    \\
    \cmidrule{1-1}
    \cmidrule{3-4}
    \cmidrule{6-7}
    \cmidrule{9-10}
    Flash &  & $5.3$ & $0.04$ &  & $9.9$ & $0.13$ &  & $24.2$ & $0.32$ 
    \\
    SAR &  & $19.0$ & $0.43$ &  & $30.1$ & $0.76$ &  & $35.8$ & $1.03$ 
    \\
    \bottomrule

\end{tabular}
\begin{tablenotes}
\item[] $^\dagger$Area (\unit{\square\milli\meter}). $^*$Power (\unit{\milli\watt}) at $0.6$V and $200$ms latency.
\end{tablenotes}
\end{threeparttable}\vspace{-2ex}

\end{table}

%% file: approx.tex
\begin{figure}[t!]
    \centering
    \includegraphics[width=\linewidth]{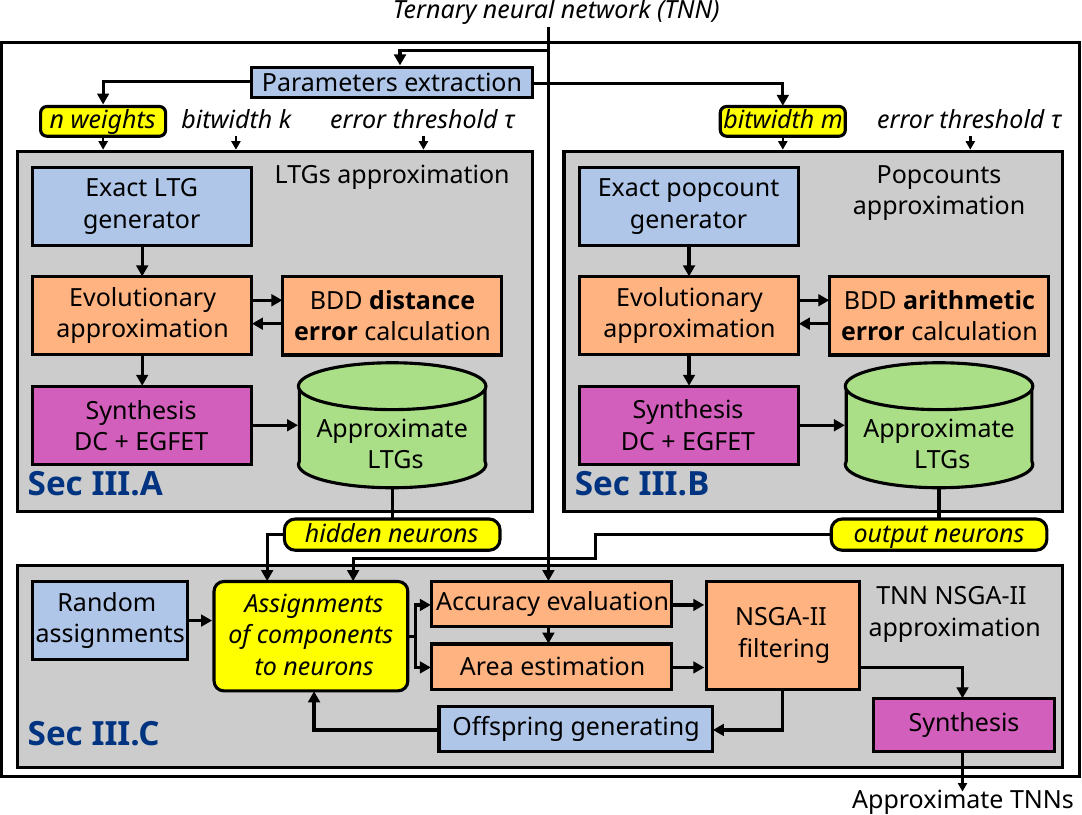}
    \caption{Overview of the proposed two-phase TNN approximation framework, including Cartesian genetic programming for popcount and LTG approximations in phase one, and their integration into bespoke TNN circuits using the NSGA-II evolutionary algorithm in phase two. The underlying technology assumed is EGFET.}
    \label{fig:overall}
\end{figure}

\section{Holistic TNN Approximation} \label{sec:axtnn}

Given a dataset, we train a TNN for each considered input precision ($1$-$4$ bits), as different precisions can result in varying TNN sizes (i.e., number of hidden neurons) as well as different sensitivity to hardware approximation.
For each trained TNN, we generate its exact bespoke circuit, and to maximize area efficiency, \textit{we approximate all its major components}.

Our proposed approximation methodology, illustrated in Fig.~\ref{fig:overall}, consists of two main phases.
We first use Cartesian Genetic Programming (CGP)~\cite{mrazek:iccad:2016} to evolve circuits that approximate the respective LTG and popcount functions, and form our approximation library of area-error Pareto-optimal approximate units for the hidden and output neurons.
In the second phase, the \blue{Non-dominated Sorting Genetic Algorithm (NSGA-II)~\cite{deb:2002}} is applied to integrate these approximate components into a bespoke TNN circuit, optimizing for maximum resource efficiency with minimal impact on accuracy. 
Factoring in the analog interfacing cost for each precision, we conduct a Pareto analysis to identify optimal designs that combine input precision and appropriate approximation.
This co-design of the analog front-end and digital classifier allows us to achieve maximum area efficiency for each target accuracy constraint.

Specifically, we replace the LTGs in the hidden layer with our approximate ones and also substitute the popcount units in the output layer with their approximate counterparts.
In bespoke circuits, where constant parameters are hardcoded, addition with a fixed term (i.e., $Z$ in~\eqref{eq:outputneuron2}) of just a few bits
is fairly cheap, making popcount the most resource-intensive part of the output neurons.
Approximate popcount units can be used across different neurons or TNNs as long as they require the same number of inputs.
Though, some output neurons may require popcount units of the same size but different $Z$ values.
Since the cost of adding $Z$ is negligible, generating approximate popcount units for every $Z$ and input size pair would unnecessarily increase our framework's complexity.
Lastly, we do not approximate \texttt{argmax}, as errors this close to the output are difficult to recover from, while the respective area gains are limited~\cite{Afentaki:ICCAD2023:axmac}.

Despite significant research on arithmetic circuits~\cite{Jiang:JPROC2020}, research on approximate LTG units is limited, as the state of the art mainly approximates the multipliers or adders of ML circuits~\cite{Armeniakos:CSUR2022}.
This is attributed to the fact that silicon-based implementations do not allow for per-neuron circuit optimizations.
Additionally, the binary output of LTGs makes generating and evaluating approximate LTGs more challenging compared to adders or multipliers with multi-bit outputs.
Similarly, despite extensive research on approximate adders, dedicated approximations for popcount circuits remain scarce.

\subsection{Approximate LTG Circuit Design} \label{subsec:axltg}
\subsubsection{LTG Netlist Approximation} 
Approximating LTGs as a single function is highly efficient, as it enables extensive optimization.  
As derived from~\eqref{eq:ltg}, the key requirement is determining whether the multi-operand sum generates a borrow. 
Thus, approximating the entire LTG in one-shot achieves superior area-error trade-offs compared to approximating and combining its components independently, as usually done in the state of the art for approximating dataflows.
An LTG configuration is defined by the number of inputs to add, subtract, and the input precision.

We employ the evolutionary Cartesian Genetic Programming (CGP) algorithm, proven effective for generating approximate circuits~\cite{mrazek:iccad:2016}.
Starting with a gate-level implementation of the exact LTG circuit, our algorithm modifies its structure to produce approximate variants, ensuring the error stays acceptable, i.e., below a predefined threshold $\tau$,
\blue{fixed to empirically set values (see Section~\ref{subsec:eval:ax_popcount} for more information on the explored range of $\tau$).}\label{rev:r3c16}
The circuit is encoded as an integer-based netlist, with mutation as the sole genetic operator.
Each mutation randomly changes a chosen integer, which can affect the connection of a gate's inputs, the gate function, or its output connection.

Operating on a population of $\lambda+1$ candidates, the initial population includes the exact circuit and $\lambda$ mutated versions.
A fitness value $F(c)$, reflecting how well each candidate approximates the target circuit while adhering to the error threshold, guides the search.
Through generations, the algorithm evolves the population, by selecting the fittest candidates and exploring/refining the solution space via mutation.
$F(c)$ represents the area of the approximate LTG circuit when implemented as a printed circuit, and is set to $\infty$ if the accuracy threshold ($\tau$) is exceeded, discarding that solution:
\begin{equation} \label{eq:errortau}  
    F(c) = \begin{cases}
        area(c), & \text{if}\, \varepsilon(c) \leq \tau \\
        \infty, & otherwise.
    \end{cases}
\end{equation}
\blue{where $\varepsilon(c)$ denotes the error of the candidate approximate LTG $c$.}\label{rev:r3c17}
To avoid costly circuit synthesis, $F(c)$ is set as the sum of the area of the netlist's gates, obtained from the EGFET PDK.
Each member $c$ in the population receives a fitness score $F(c)$, with the highest-scoring individual becoming the parent of the next population.
This parent then generates $\lambda$ new candidate solutions through mutation.
The process terminates when either the number of iterations exceeds a set maximum, or a predefined time limit is reached.

\subsubsection{LTG Error Analysis}
LTGs are basically relational operators with Boolean output, where the output is determined by comparing a weighted sum with zero; hence, Hamming-based error metrics are unsuitable~\cite{mrazek:iscas:2025:axmed}.
Given a set of weights $\boldsymbol{w}$, the exact LTG ($\mathrm{LTG}_{\boldsymbol{w}}$) and its approximate variant ($\mathrm{\hat{LTG}}_{\boldsymbol{w}}$), we adopt a more appropriate distance error metric for such functions: 
\begin{equation}
    \label{eq:distance}
D(\boldsymbol{x}) =\begin{cases}
			0,\quad  \text{if}\,\, \mathrm{LTG}_{\boldsymbol{w}}(\boldsymbol{x}) = \mathrm{\hat{LTG}}_{\boldsymbol{w}}(\boldsymbol{x})\\  
              |\sum_{i=0}^{n} w_i x_i |\!+\!1,\quad  \text{otherwise}.
		 \end{cases}
\end{equation}
\blue{If $\mathrm{LTG}_{\boldsymbol{w}}$ and $\mathrm{\hat{LTG}}_{\boldsymbol{w}}$ differ, it means that the two functions produced outputs of opposite signs--that is, one computed a positive weighted sum and the other a negative one.  
However, since $\mathrm{\hat{LTG}}_{\boldsymbol{w}}$ is implemented as a black-box function via CGP, we only observe its binary output ($0$ or $1$), without any direct insight into how closely it approximated the actual weighted sum.  
To capture the severity of such mismatches, we define an error metric $D(\boldsymbol{x})$ that quantifies the difference/distance between the weighted sum of $\mathrm{LTG}_{\boldsymbol{w}}$ and the decision threshold at $0$, using the exact LTG as the reference value.
This provides a meaningful measure of how far the approximate function was from producing the correct output, assuming that $\mathrm{\hat{LTG}}_{\boldsymbol{w}}$ flipped the sign of the sum.  
An offset of $1$ is added to handle cases where the sum is zero, but $\mathrm{\hat{LTG}}_{\boldsymbol{w}}$ wrongly produced a negative sign.
}\label{rev:r3c18-r3c19}


Using $D(\boldsymbol{x})$ we define three statistical error measures, i.e., error probability $\varepsilon_{ep}$, mean distance error $\varepsilon_{mde}$, worst-case distance error $\varepsilon_{wcde}$, and normalized distance error $\varepsilon_{epmde}$ involving the erroneous outputs only:
\begin{equation}
    \begin{gathered}
    \varepsilon_{ep} = \frac{1}{K} \sum_{\forall \boldsymbol{x}} \llbracket D(\boldsymbol{x}) \neq 0 \rrbracket,\quad 
    \varepsilon_{mde} = \frac{1}{K} \sum_{\forall \boldsymbol{x}} D(\boldsymbol{x}), \\
    \varepsilon_{wcde} = \max_{\forall \boldsymbol{x}} D(\boldsymbol{x}),\quad
    \varepsilon_{epmde} = \frac{\varepsilon_{mde}}{\varepsilon_{ep}}
\end{gathered}\label{eq:errormetrics}
\end{equation}
where $K$ represents the size of the input domain for $\boldsymbol{x}$ and $\llbracket \cdot \rrbracket$ denotes the Iverson bracket.


\begin{figure}[t!]
    \centering 
    \includegraphics[width=0.9\columnwidth]{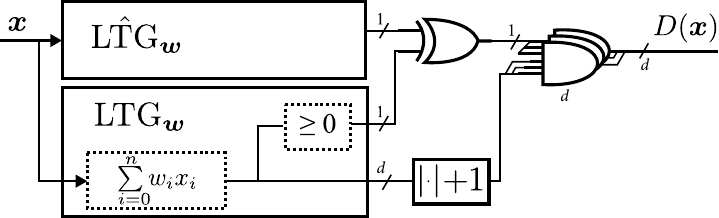}
    \caption{Miter circuit for calculating distance error $D(\boldsymbol{x})$.}
    \label{fig:miter}
\end{figure}

For $k$-bit inputs ($x_i$), evaluating the full input space involves $K=2^{nk}$ stimuli.
To expedite error evaluation, Binary Decision Diagrams (BDDs)~\cite{mrazek:isvlsi:2022} with a miter circuit (Fig.~\ref{fig:miter}) are used to efficiently compute $\varepsilon_{ep}$, $\varepsilon_{mde}$, and $\varepsilon_{wcde}$.

\subsection{Approximate Popcount Circuits}
\label{subsec:ax_popcount}
A popcount configuration is defined by the number of inputs to count.
An $m$-bit popcount operation sums $m$ binary inputs, typically implemented using a tree structure of adders.
To approximate this exact circuit, we also employ the evolutionary CGP algorithm.
Similarly, the searching algorithm's optimization criterion is the estimated area for the target technology (e.g., EGFET PDK~\cite{Bleier:ISCA:2020:printedmicro}), while maintaining the error $\varepsilon$ below a specific threshold, consistent with previous error-oriented approximation research~\cite{mrazek:iccad:2016}.
This error constraint $\tau$ is again incorporated into the fitness function as in \eqref{eq:errortau}.

Popcount circuits produce standard arithmetic outputs, allowing the use of conventional error metrics in the approximation process: mean arithmetic error $\varepsilon_{mae}$ and worst-case arithmetic error $\varepsilon_{wcae}$.
These metrics calculate the average or maximum error across all input combinations.
Assuming $P(\boldsymbol{p})$  is the output of the exact popcount for a set of $m$ binary inputs $\boldsymbol{p}$ and $\hat{P}(\boldsymbol{p})$ is its approximate counterpart, $\varepsilon_{mae}$ and $\varepsilon_{wcae}$ are calculated as follows:
\begin{equation}
    \begin{gathered}
    \varepsilon_{mae} = \frac{1}{2^m} \sum_{\forall \boldsymbol{p}} |P(\boldsymbol{p})-\hat{P}(\boldsymbol{p})|, \\
    \varepsilon_{wcae} = \max_{\forall \boldsymbol{p}}  |P(\boldsymbol{p})-\hat{P}(\boldsymbol{p})|
\end{gathered}\label{eq:pcerrormetrics}
\end{equation}
However, for a large number of inputs $m$ in popcount circuits, evaluating all $2^m$ input vectors becomes impractical.
Hence, BDDs~\cite{mrazek:isvlsi:2022} are also used in this case for error assessment.

\subsection{Approximate TNN Design}
\subsubsection{Multi-objective optimization} Our framework aims to select the optimal approximate unit for each neuron in the exact TNN, i.e., to identify the most suitable replacement for each exact LTG and popcount from the approximate units in our library.
If a required LTG or popcount configuration is not already available in our library, we generate the respective approximate units offline as a one-time effort.

We transform the approximation selection into a multi-objective optimization and solve it using the NSGA-II algorithm~\cite{deb:2002}, with objectives to minimize accuracy loss and area estimation.
In printed electronics, leakage power dominates power consumption.
Hence, minimizing area minimizes power as well.
We encode our optimization problem as an integer list, where each integer selects a component from the library for the respective neuron.
All components, including the TNN, (approximate) LTGs, and (approximate) popcounts, are described in both Verilog and Python, enabling efficient accuracy estimation in Python.
Similarly, area estimation is conducted in Python using our surrogate area model.
The Pareto-optimal circuits identified by NSGA-II are synthesized and evaluated using \blue{Electronic Design Automation (EDA)} tools to determine the actual area and power values of our approximate TNNs.

\subsubsection{Area Estimator}
Since area is additive, 
we use as a surrogate area model the sum of the areas of the selected approximate units.
During library generation, all approximate units are synthesized to determine their area.
Implementing a holistic approximation that covers all major TNN components enables our estimator to achieve high accuracy.
The only non-approximated components are \texttt{argmax} and constant addition by $Z$, which are insignificant in size.

We validate our estimator across $500$ approximate TNNs with various datasets and approximation levels.
It demonstrates a \textit{Pearson correlation of $0.995$ and an R$^2$ value of $0.969$}, indicating both near-perfect linear correlation, enabling us to effectively guide our optimization toward more area-efficient solutions, but also precise area estimation.

%% file: experimental.tex
\section{Results and Analysis}\label{sec:experimental}
The goal of the presented research is to develop arbitrary input precision bespoke TNN circuit designs that maximize hardware efficiency while minimizing any decrease in accuracy.
Initially, we evaluate the proposed hardware implementations and the efficacy of our approximation methods.
Subsequently, we conduct a comparative analysis against the current state-of-the-art printed neural networks.

\textbf{Experimental Setup:}
For our evaluation, we use eight datasets from the UCI ML repository~\cite{Dua:2019:UCIdatasets}, selected for two reasons: they allow direct comparison with the state of the art~\cite{Mubarik:MICRO:2020:printedml,Afentaki:DATE2024:gatrain, Afentaki:ICCAD2023:axmac,Armeniakos:TC2023:codesign, mrazek:iccad:2024}, and are well-suited for sensor-based printed applications as detailed in~\cite{Mubarik:MICRO:2020:printedml,Weller:2021:printed_stoch}.
Such neural networks are orders of magnitude simpler than contemporary Convolutional Neural Networks and Large Language Models targeted by silicon systems.  
\textit{However, printed circuits can integrate only a few hundred to thousand gates per chip, making them both representative and highly complex examples for printed electronics to address.}  
For instance, Pendigits is the most complex dataset explored by the current state of the art~\cite{Mubarik:MICRO:2020:printedml,Weller:2021:printed_stoch,Armeniakos:DATE2022:axml,Armeniakos:TC2023:codesign,Afentaki:ICCAD2023:axmac,Afentaki:DATE2024:gatrain,mrazek:iccad:2024}, yet no implementation to date has achieved an adequate trade-off in terms of area, power, and accuracy.
We use Synopsys Design Compiler, PrimeTime, and the EGFET standard cell library~\cite{Bleier:ISCA:2020:printedmicro} for synthesis and hardware evaluation.
The voltage supply for all our circuits is set to 0.6~V, in line with EGFET capabilities and consistent with prior work in the field~\cite{Afentaki:ICCAD2023:axmac,Afentaki:DATE2024:gatrain}.
The Verilog description of our TNNs is generated directly from the Python model using the description of the LTGs and popcounts in our library.

\textbf{TNN Training:}
Exact TNN models are used as the baseline for accuracy comparisons.
Input features are normalized within $[0, 1]$, datasets follow a $70\%/30\%$ split for training/testing and TNNs are trained using QKeras~\cite{Coelho:arXiv:2020:qkeras} and its \texttt{quantized\_bits} functionality.
We employ the Adam optimizer for up to $30$ epochs of training (with early stopping). 
The learning rate for each network is obtained through Bayesian optimization, aiming to maximize the model's inference accuracy with a maximum of $100$ attempts.
Learning rate parameters are selected from the range of $0.001$ to $0.01$.
Hidden neurons are selected via grid search within the $[1, 50]$ range.
For each dataset, we train a TNN for input precisions from $1$ to $4$ bits and, for each precision, select the TNN with the smallest hidden-layer size that achieves accuracy closest to the exact MLP~\cite{Mubarik:MICRO:2020:printedml}.

\subsection{Approximate Components Evaluation}
\label{subsec:eval:ax_components}
\subsubsection{Popcount Circuits}
\label{subsec:eval:ax_popcount}
We generate approximate popcount circuits with the following constraints and configurations.
The error limits $\tau_{mae}$ and $\tau_{wcae}$ are logarithmically distributed from $0.1$ to $0.5\cdot2^g$ and from $1$ to $0.5\cdot2^m$, with $g = \lceil \log_2 m \rceil$, resulting in $2,\!090$ approximate circuits.
We also set CGP search termination criteria to $30$, $60$, and $300$ minutes for popcount sizes $m<16$, $m<32$, and $m<60$, respectively.
Using BDD evaluation, we achieve average speeds of $7,\!759$ and $1,\!362$ evaluations per second for $16$-bit and $32$-bit popcounts.
For $60$-bit popcount circuits, this number falls to $25$ evaluations per second.
To account for this, we extend the time limit, thereby increasing the chances of identifying solutions with better area-accuracy trade-offs.

\begin{figure}[t]
    \centering
    \includegraphics[width=0.9\columnwidth]{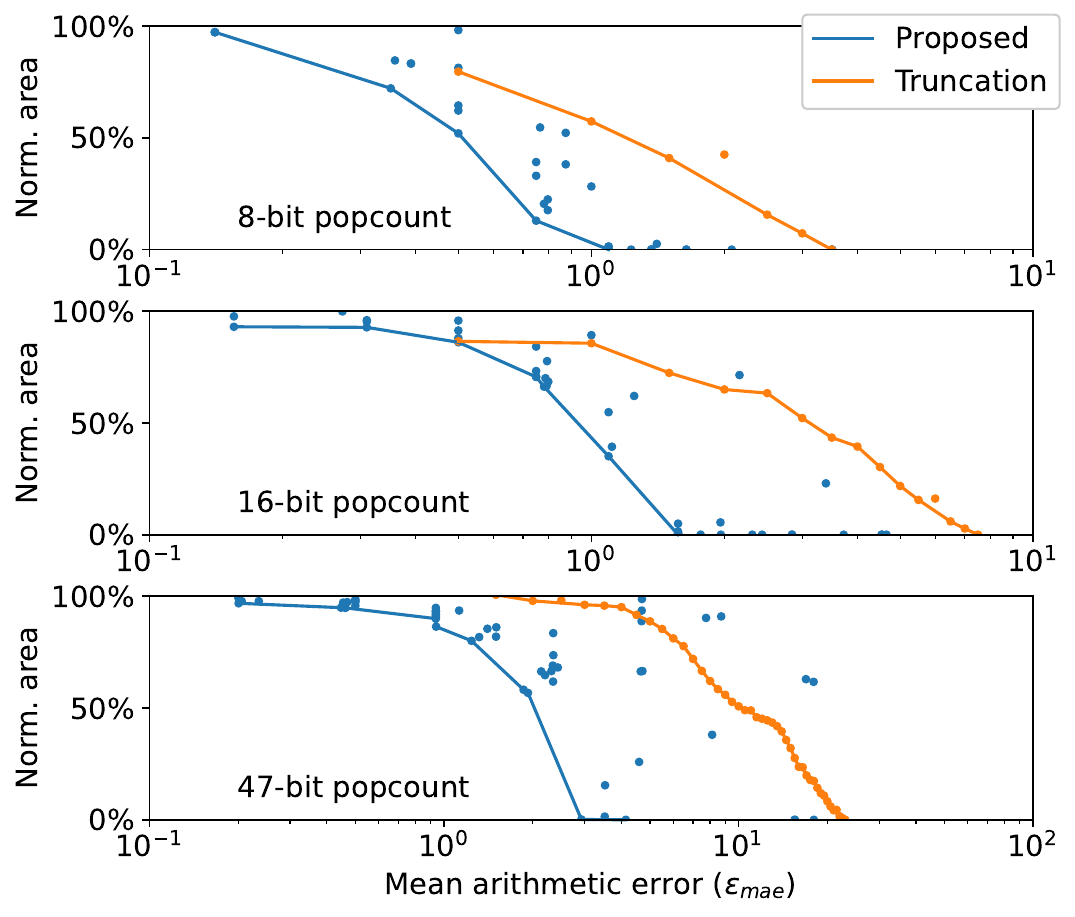}
    \caption{Comparison between the proposed approximation approach and the previously used truncation technique for popcount circuits of various sizes. The results displayed are based on post-synthesis area measurements. 
    }
    \label{fig:pc_selected}
\end{figure}

A visualization of the obtained results for three popcount circuit sizes is provided in Fig.~\ref{fig:pc_selected}.
The featured circuits are synthesized and evaluated based on the error metrics and methodology described above.
For easier comparison, area results are normalized relative to the exact popcount circuit of the same size.
Our results indicate that as the error limit increases, CGP optimization identifies increasingly area-efficient solutions.
Additionally, we assess the efficacy of our approximation approach through a comparison of our popcount circuits with variants derived using the truncation approximation approach, which has previously been applied in a wide range of approximate circuits, including MLPs~\cite{venkataramani:axnn}.
As shown \blue{in Fig.~\ref{fig:pc_selected}}\label{rev:r3c21}, our approximation achieves significantly superior trade-offs compared to truncation, and delivers about $2$x area reduction for $\varepsilon_{mae}$ of only $0.5$, $1.1$, and $1.9$ for $8$-bit, $16$-bit, and $47$-bit popcounts, respectively.
Similar results to those shown in Fig.~\ref{fig:pc_selected} are observed for all popcount circuit sizes.

\begin{figure}[t!]
\centering
    \includegraphics[width=\columnwidth]{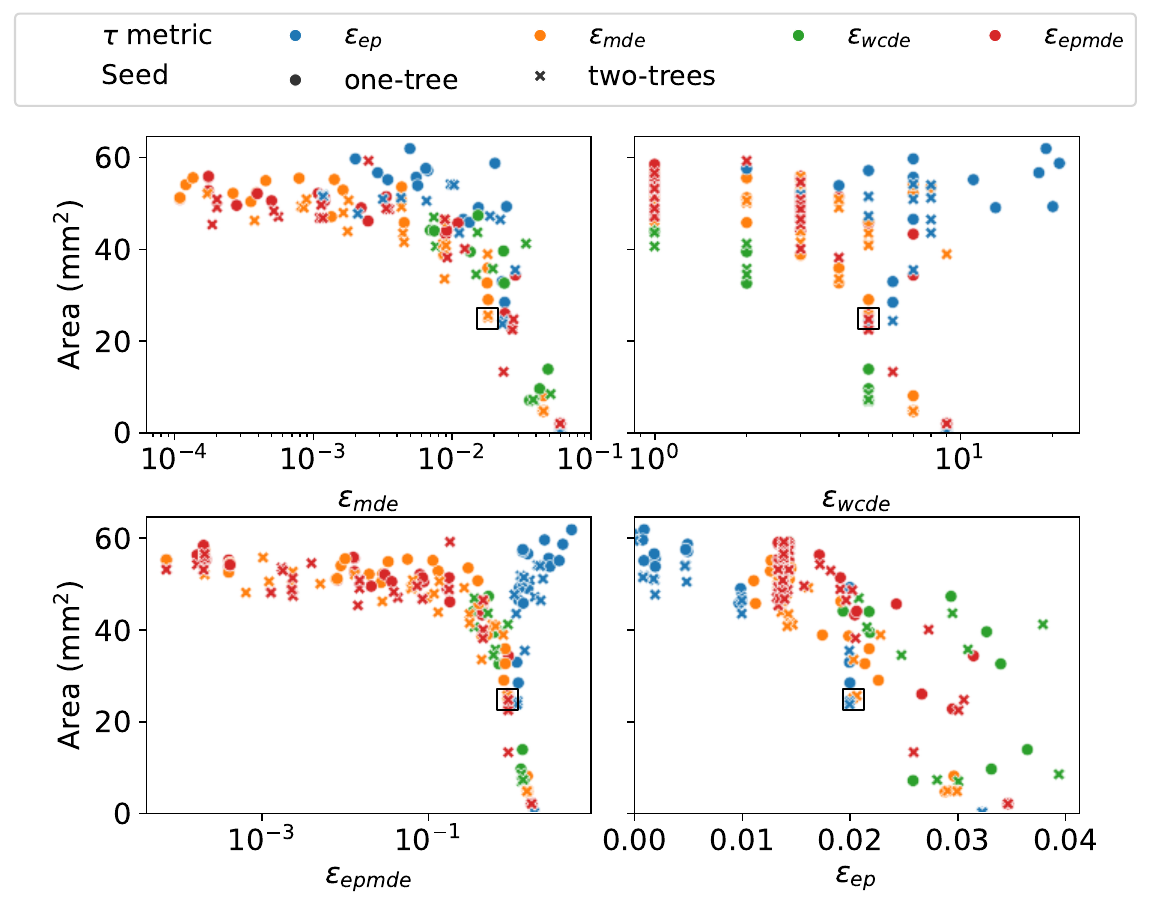}
    \caption{Approximate LTGs realizing Pendigits' first hidden neuron with ten 2-bit inputs. Exact LTG area: 59.1mm$^2$. The results displayed are based on post-synthesis area measurements.
    \blue{Highlighted design points are in rectangles.}}
    \label{fig:neuron}
\end{figure}

\subsubsection{LTG Circuits}
\begin{figure}[t]
    \centering
    \includegraphics[width=0.9\columnwidth]{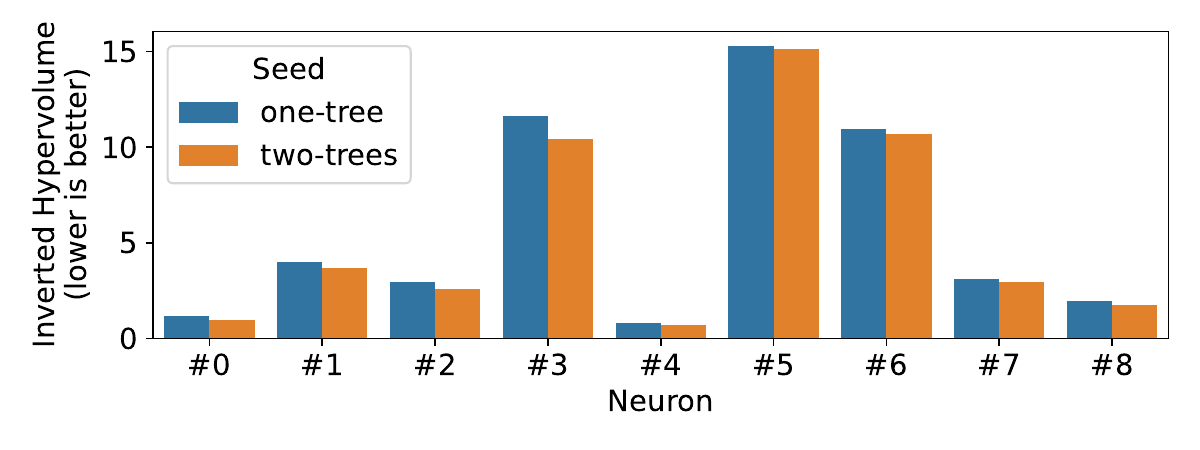}\vspace{-2ex}
    \caption{\blue{Inverted hypervolume (i.e., AUC from [0, 0] point) for 2-bit LTGs realizing the hidden neurons of our Pendigits TNN.
    Both implementations from Fig.~\ref{fig:neuron} (i.e., one adder tree or two adder tree architecture) are considered.}}
    \label{fig:neurons2b_hypervolume}
\end{figure}

The exact LTG can be implemented either as one adder tree that sums or subtracts inputs and checks for a borrow, or as two separate adder trees---one to sum the inputs to be added and another to sum the inputs to be subtracted---whose results are then subtracted.
We generate our approximate LTGs (Section~\ref{subsec:axltg}) for these exact implementations and optimize them against the error metrics of~\eqref{eq:errormetrics}.
Fig.~\ref{fig:neuron} depicts the error-area trade-off for all our approximate LTGs, using a ten $2$-bit inputs ($n$=10, $k$=2) neuron as an example.
As shown, the implementation with two adder trees provides more efficient solutions ({\footnotesize\faRemove} are mainly below $\bullet$ for similar error).
\blue{This is also confirmed by the inverted hypervolume (i.e., area under the curve (AUC), where lower values indicate better performance) for the same neurons, presented in Fig.~\ref{fig:neurons2b_hypervolume}.
On average, the two adder tree implementation achieves an $8.43 \pm 4.74\%$ improvement over the single case, and is therefore used next.}\label{rev:r3c6}
Moreover, Fig.~\ref{fig:neuron} provides a descriptive enough illustration of the gains of our LTG approximation.
For example, for $\varepsilon_{mde}\leq 0.018$ in Fig.~\ref{fig:neuron}, $58\%$ area reduction is achieved compared to the exact LTG \blue{(indicated by black rectangles in all subplots).}\label{rev:r3c22}
Similar results are obtained for all LTG configurations required by other datasets and neurons.

For the TNNs examined, the LTGs required are relatively large, with $n \cdot k$ averaging $22$.
When $n \cdot k = 32$ and $40$, the BDD-based LTG error evaluation takes at most $0.09$s and $0.21$s, respectively, which is $450$x and $1850$x faster than vectorized simulation~\cite{Sekanina2019}, requiring $42$s and $392$s.  
This rapid error assessment enables efficient traversal of the approximation space and exploration of error-area optimal LTGs.

\subsection{Approximate TNN Evaluation}
\label{sec:eval:approx_tnn}
The final stage of our evolutionary approximation approach utilizes the NSGA-II algorithm to construct a TNN design.
This design incorporates components from our LTG (hidden neurons) and popcount (output neurons) circuit libraries.
We apply our proposed approximation to all examined datasets.  
For each TNN, if the required LTG and popcount configurations are not already in our library, they are generated as described above.

\begin{figure}[t]
    \centering
    \includegraphics[width=\columnwidth]{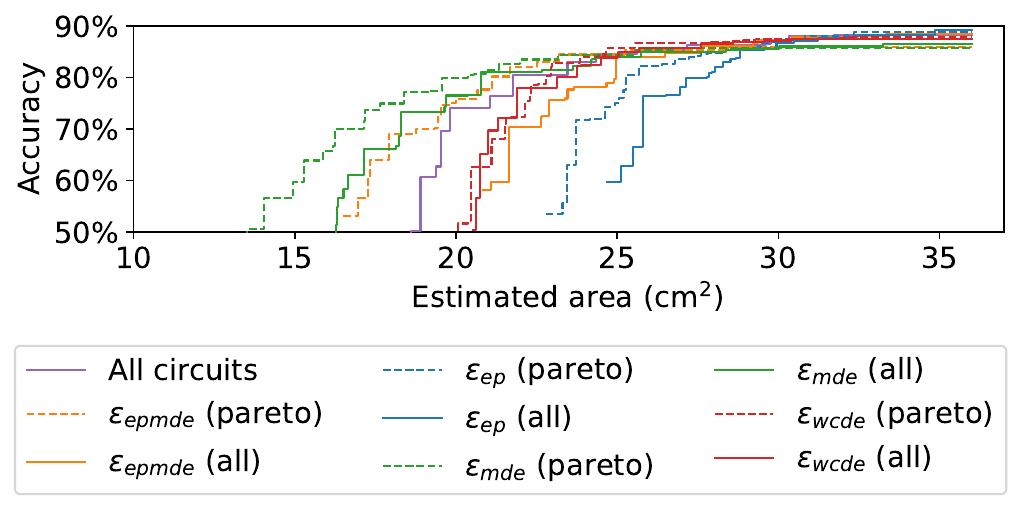}\vspace{-1em}
    \caption{Approximate 2-bit Pendigits TNNs using different LTG libraries, each optimized against specific error metric.}
    \label{fig:moopn2}
\end{figure}

\begin{figure}[t]
    \centering
    \includegraphics[width=\columnwidth]{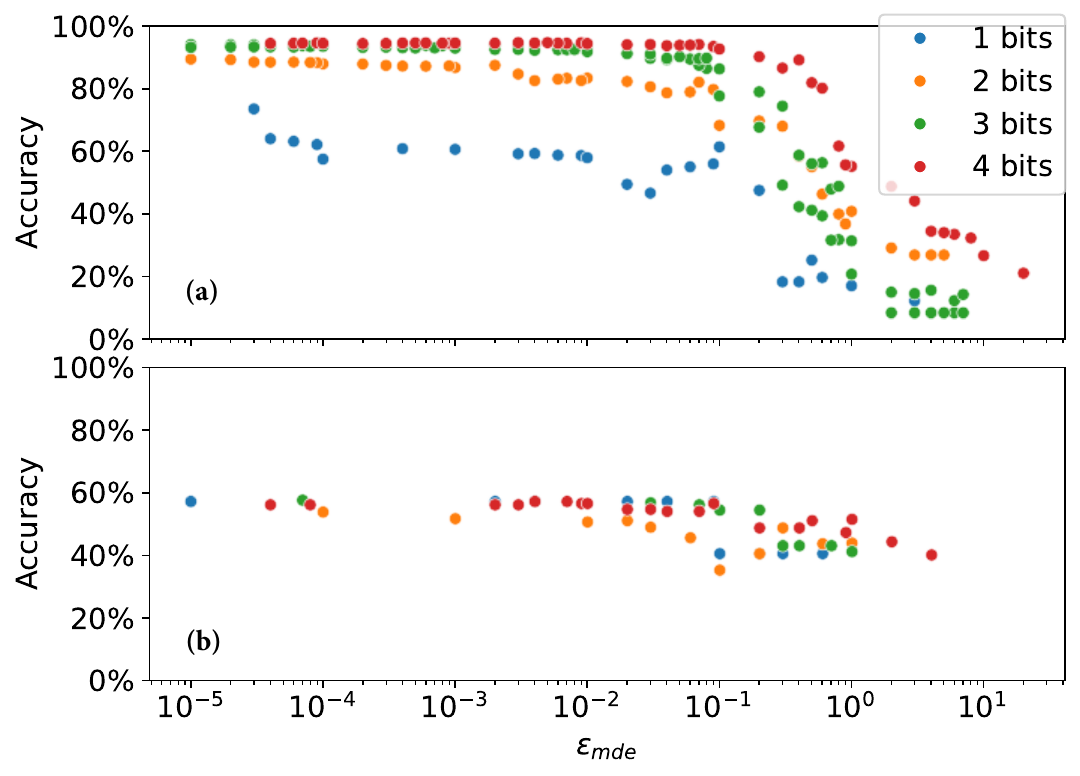}\vspace{-1em}
    \caption{\blue{Correlation analysis between our distance error metric ($\varepsilon_{mde}$) and classification accuracy for (a) Pendigits and (b) RedWine TNNs, when approximate LTGs of the same $\varepsilon_{mde}$ error are used for their hidden neurons.}}
    \label{fig:emde_correlation}
\end{figure}

Fig.~\ref{fig:moopn2} shows the results of our multi-objective optimization for the $2$-bit Pendigits dataset using various approximate LTG libraries, each evolved with a distinct error metric.
Libraries containing only Pareto-optimal LTG units (``pareto'') yield superior outcomes.
For high accuracy, LTG designs optimized for the $\varepsilon_{wcde}$ metric are better, while those evolved for the $\varepsilon_{mde}$ metric are preferable for maximizing area gains.
Similar findings to Fig.~\ref{fig:moopn2} apply to other datasets.
Hence, for each TNN, the required LTG circuits are approximated against both $\varepsilon_{mde}$ and $\varepsilon_{wced}$ to populate our library.
\blue{In Fig.~\ref{fig:emde_correlation}, we further investigate the role of our $D$ error metric in approximating LTG units for achieving area-efficient and high-accuracy approximate TNNs.
Fig.~\ref{fig:emde_correlation} presents a correlation analysis between $\varepsilon_{mde}$ and classification accuracy, when all TNN hidden neurons are replaced with approximate LTGs of identical $\varepsilon_{mde}$, for a wide range of error values.
Two TNNs are used in this analysis: our most complex one (Pendigits) and a relatively small one (RedWine).
A strong non-linear correlation between our $\varepsilon_{mde}$ metric and the classification accuracy can be observed, especially as the network's knowledge capacity increases--i.e., when moving from smaller (RedWine) to larger (Pendigits) networks, or from lower ($1$-bit) to higher ($4$-bit) input precision.
Fig.~\ref{fig:emde_correlation} validates that we obtain the expected correlation between neural network inference accuracy and approximate arithmetic units, and confirms that our $\varepsilon_{mde}$ metric is suitable for effectively guiding approximation exploration toward high-accuracy regions.
Specifically, as $\varepsilon_{mde}$ (and thus area savings) increases, there exists a broad region where minimal accuracy degradation is observed at the TNN level, maximizing, thus, the area efficiency of our approximate TNNs.
}\label{rev:r2c3}

\begin{figure}[t]
    \centering
    \includegraphics[width=\columnwidth]{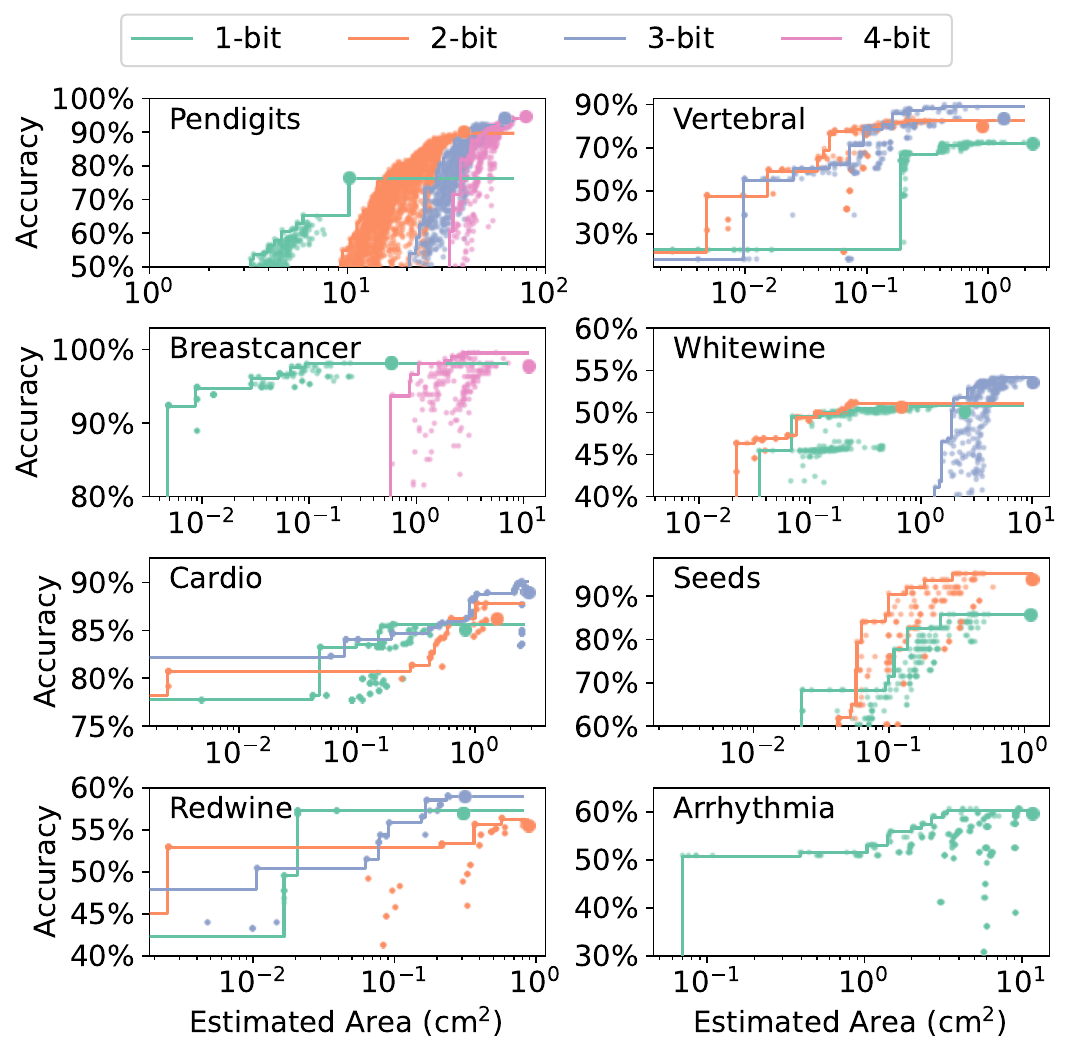}
    \caption{Accuracy-area analysis of approximate TNNs generated by our framework. X-axis shows the estimated area, with big circles indicating the respective exact TNNs for each precision.}
    \label{fig:mooall}
\end{figure}

Fig.~\ref{fig:mooall} shows the accuracy-area trade-off for all explored solutions during our multi-objective optimization, with the Pareto fronts (per input precision) annotated by lines.
As Fig.~\ref{fig:mooall} includes $15,\!000$ designs, synthesizing all is impractical, so we report the estimated area using our precise surrogate model.
Post-synthesis results for key points are presented in Table~\ref{tab:comp}.
Note that the area estimation in Fig.~\ref{fig:mooall} reflects only the TNN classifier area, excluding any interfacing costs.
In Fig.~\ref{fig:mooall}, we include only the input precisions that result in at least one Pareto-optimal point for each dataset,
\blue{since precisions that lead to only dominated solutions (i.e., with higher area and lower accuracy) are suboptimal in our evolutionary-based optimization.}\label{rev:r3c23}
As shown, our framework consistently delivers a smooth accuracy-area trade-off.
Supporting arbitrary input precision allows us to effectively populate the accuracy-area Pareto front, as it is primarily composed of designs with varying precisions, with only one precision, especially the low cost $1$ bit, rarely dominating.

Two cases require further discussion. 
For Pendigits with $1$-bit input precision, our framework maintains a smooth accuracy-area trade-off despite a significant initial accuracy loss w.r.t the exact $1$-bit TNN.
Pendigits, being the most complex dataset examined, becomes highly sensitive to approximation with only $1$-bit inputs.
This is evident from the $1$-bit exact Pendigits TNN achieving just $76$\% accuracy, while the target MLP reaches $94$\%~\cite{Mubarik:MICRO:2020:printedml}.
This further highlights that it is mandatory to support arbitrary input precision, as $1$-bit precision only (e.g.,~\cite{mrazek:iccad:2024}) may be insufficient in many cases.

For Arrhythmia, we present results only for $1$-bit precision, even though the $1$-bit exact TNN exhibits a $2$\% accuracy loss compared to the respective exact MLP, while the $2$-bit exact TNN features no accuracy loss.
Our framework uses CGP to evolve exact LTG circuits into approximate variants and supports exact design with up to $90$ input bits.
The $2$-bit Arrhythmia LTG circuits feature over $100$ inputs of $2$ bits, totaling more than $200$ input bits, and our CGP algorithm could not find suitable approximations within a reasonable timeframe.
However, considering datasets like Arrhythmia, with more than $250$ input features, might be unrealistic for printed applications, particularly when factoring in ADC costs.
Nevertheless, our framework still provides a strong trade-off, delivering $86$x lower area than the printed exact MLP~\cite{Mubarik:MICRO:2020:printedml}, for only $5$\% accuracy loss (see Table~\ref{tab:comp}).
In addition, our framework can be combined with pruning---which is not considered in this work to ensure fairness in comparisons---to reduce inputs to LTGs and potentially mitigate such issues.
\blue{Alternatively, we can follow conventional hierarchical approaches for the generation of approximate LTGs with more than $200$ input bits.
This involves breaking the LTG down to components (two or more adders and one comparator) and following a typical procedure for approximating dataflows~\cite{Mrazek2019AutoAx,Zervakis:TCASII2019}, e.g., creating a library of approximate components and then combining them through multi-objective optimization (e.g., NSGA-II). 
Such approaches might lead to some optimality loss but can address scalability issues in LTG approximation.}\label{rev:r2c2}

\input{tab_soa_cmp}

\subsection{Comparison with the state of the art}
\label{subsec:comp_soa}
Table~\ref{tab:comp} compares our approximate TNNs with the current state-of-the-art digital printed neural networks.
We specifically evaluate against the approximate TNNs in~\cite{mrazek:iccad:2024}, which use $1$-bit inputs and require only ABCs, the approximate MLPs from~\cite{Armeniakos:TC2023:codesign}, which employ a conservative approximate $8$-bit multiplication scheme, and the approximate MLPs from~\cite{Afentaki:DATE2024:gatrain,Afentaki:ICCAD2023:axmac}, which apply aggressive approximation using power-of-2 weights.
~\cite{Armeniakos:TC2023:codesign,Afentaki:DATE2024:gatrain,Afentaki:ICCAD2023:axmac} also approximate the additions and use $4$-bit inputs to enable high accuracy, despite the applied approximations.
In Table~\ref{tab:comp}, we consider two accuracy loss thresholds relative to the baseline exact printed MLP~\cite{Mubarik:MICRO:2020:printedml}: $2\%$ for high accuracy and $5\%$ for high hardware gains.
For improved readability, for each dataset and accuracy threshold, we report only the most area-efficient approximate printed MLP~\cite{Armeniakos:TC2023:codesign,Afentaki:DATE2024:gatrain,Afentaki:ICCAD2023:axmac}.
The table also includes the input precision used by each neural network, emphasizing the significance of our framework's support for arbitrary input precision.
As shown in Table~\ref{tab:comp}, different datasets and accuracy thresholds mandate varying precision for our TNNs.
The reported area and power values include the analog-to-digital interface cost, which our approach minimizes by selecting the most area-efficient solution with respect to both digital classifier and interfacing costs.
This often involves choosing the minimum necessary precision due to high ADC overheads.
For consistency, the interfacing costs reported in Section~\ref{subsec:adc} are used in all designs in Table~\ref{tab:comp}.
Hereafter, references to ADC or interfacing costs refer to both ADC and/or ABC.

As shown in Table~\ref{tab:comp}, with a maximum $2\%$ accuracy loss compared to the respective exact TNN, our approximate ones achieve on average $2.9$x and $4.7$x lower area and power, including ADC costs.
These values increase to $10.2$x and $11.1$x when excluding ADC costs, i.e., with respect to only the TNN classifier, demonstrating the high efficiency of our approximation.
Compared to the baseline exact MLPs~\cite{Mubarik:MICRO:2020:printedml}, our TNNs feature $88$x and $783$x, on average, lower area and power (including ADC costs) for up to $2\%$ accuracy loss.

Compared to the most efficient approximate MLP~\cite{Armeniakos:TC2023:codesign,Afentaki:DATE2024:gatrain,Afentaki:ICCAD2023:axmac} in each case, our approximate TNNs achieve, on average, $21$x lower area and $67$x lower power for the $2$\% accuracy loss threshold.
For the $5$\% threshold, these gains increase to $36$x and $139$x, highlighting the limited scalability of existing printed MLP approximation frameworks.
Excluding ADC costs, the latter gains become $24$x and $53$x.
This demonstrates two key points: i) our framework can apply high degree of approximation and achieve significant area efficiency, even with limited input precision; and ii) minimizing the cost of ADCs is crucial, as for half of the printed neural networks in Table~\ref{tab:comp}, including ADCs increases area by over than $47$\%.
Finally, it is noteworthy that~\cite{Afentaki:DATE2024:gatrain,Afentaki:ICCAD2023:axmac} fail to meet the $2\%$ accuracy loss constraint for Pendigits and the $5\%$ loss for Seeds and Vertebral.

The comparison with our prior work on approximate TNNs~\cite{mrazek:iccad:2024} further highlights the effectiveness of this solution.
For $1$-bit inputs (as in~\cite{mrazek:iccad:2024}) and identical accuracy, our TNNs achieve, on average, $1.8$x and $2.9$x lower area and power (including ADC costs), and $5.5$x and $5.2$x savings when considering only the TNN classifier.
These gains (especially the latter) underline the superiority of our approximation strategy.
\cite{mrazek:iccad:2024}~relies on approximate popcount units in both hidden and output layers, requiring multi-bit exact comparators in the hidden layer.
In contrast, \textit{we fully approximate each hidden neuron as a single function, improving area-accuracy trade-offs for both the hidden neurons and the TNN overall}.
Moreover, \cite{mrazek:iccad:2024} is restricted to $1$-bit inputs due to its popcount-only approximation and uses only binary weights in the output layer for implementation simplicity. 
Therefore, \cite{mrazek:iccad:2024}~fails to meet the $2\%$ constraint for Cardio and WhiteWine, and does not even achieve a $5\%$ accuracy loss for Pendigits, Seeds, and Vertebral.
Supporting arbitrary input precision and ternary weights in both layers, \textit{our framework delivers hardware-efficient solutions across all datasets and accuracy thresholds}.

Importantly, \textit{only our TNNs} meet the $30$mW power constraint across all datasets, enabling battery operation with an existing printed battery, e.g., Molex $30$mW~\cite{Mubarik:MICRO:2020:printedml}.
In printed applications, design feasibility is of paramount concern, taking precedence over maximizing classification accuracy, making the $5$\% accuracy loss a mostly acceptable constraint~\cite{Afentaki:ICCAD2023:axmac}.

\blue{Table~\ref{tab:comp} presents a comparative study against the printed neural networks state-of-the-art.
For completeness, we also measure the accuracy that could be achieved by a silicon system, e.g., a deep MLP (more than $5$ hidden layers and more than $50$ neurons per layer) that could run on a TinyML microcontroller. 
We observed that our TNNs remain in most cases within $5\%$ accuracy loss of the respective FP32 silicon-oriented MLP.
The maximum accuracy drop is observed for the Arrhythmia dataset (i.e., $16$\%), which, due to its increased number of input features, poses a priori a significant challenge for printed implementations~\cite{Mubarik:MICRO:2020:printedml}.
From this analysis, we conclude that, given the immense computational limitations of printed technologies, our TNNs deliver reasonable accuracy compared to the silicon-oriented MLPs, while maintaining realistic area requirements and power consumption compatible with existing printed batteries---thus enabling feasible implementations.
It is also noteworthy that these silicon-oriented MLPs require over $24{,}000$ MAC operations, which would translate to unrealistic area requirements in printed electronics.}\label{rev:r2c4}

\blue{Finally, we investigate the impact of our approximations on classification boundaries by analyzing the confidence margins of our approximate TNNs relative to the exact ones.
By calculating the confidence margins for our approximate circuits of Table~\ref{tab:comp}, we observe that they are always maintained above $1$ and often remain comparable to---or even better than---the exact TNN, demonstrating that classification boundaries are preserved under the applied approximations.
Note, classification boundaries can be seamlessly incorporated into our optimization, by modifying \eqref{eq:errortau} accordingly.}\label{rev:r1c2}

\begin{figure}[t]
    \centering
    \includegraphics[width=\columnwidth]{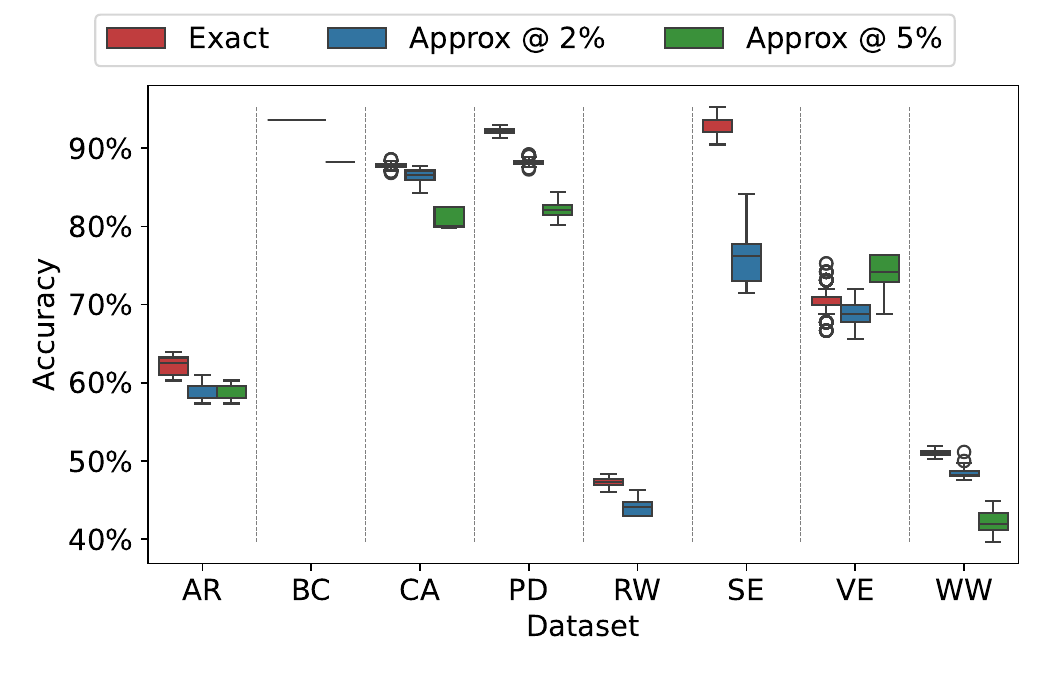}
    \caption{Classification accuracy of our TNNs in Table~\ref{tab:comp} under $10$\% process variation in the analog interfacing.}
    \label{fig:variation}
\end{figure}

\subsection{\blue{Discussion on Process Variation}}
\label{subsec:variation}
\blue{Process variation is an inherent and significant phenomenon in printed electronics, stemming from the low-resolution printing process.
Nevertheless, our TNN classifiers are purely digital and their functionality and output may only be affected by improper timing caused by process variations.
In addition, our TNN classifiers implement a fully parallel, unfolded architecture without needing any sequential elements.
Therefore, the absence of sequential logic, combined with the large clock slack we employ, makes the functionality our digital classifiers robust to timing variations caused by variability.

Nevertheless, variability may affect the required ADCs/ABCs, which, as analog components, are more susceptible to process variations and can therefore be subject to degradation under imperfections during the manufacturing process.
Variations in the analog interfacing may eventually lead to misclassifications in the digital part.
For example, variations in the resistor ladder of the Flash ADC/ABC can alter the reference voltage levels of the quantization segments, potentially causing quantization errors at the ADC/ABC output that manifest as noise (or input perturbations) at the input layer of our TNNs.
To evaluate the impact of process variation in the analog part on classification accuracy, we perform a Monte Carlo analysis assuming high variation levels (i.e., $10$\% in the analog components).
Specifically, for our exact and Pareto-optimal approximate TNNs of Table~\ref{tab:comp}, we run a 200-point Monte Carlo simulation, capturing the inference accuracy on the test dataset under process variation in the ADCs/ABCs.

Fig.~\ref{fig:variation} illustrates the obtained Monte Carlo results.
As expected, classification accuracy is influenced by process variation, since small variances in the resistor ladder can slightly shift the voltage reference levels, potentially leading to significant errors due to the quantization effect of low-precision ADCs.
Nevertheless, the accuracy variation observed in Fig.~\ref{fig:variation} remains well constrained in most cases, demonstrating the inherent robustness of TNNs to input perturbations.
For example, the average standard deviation across all TNNs in Fig.~\ref{fig:variation} is only $0.8$\%.
The largest dispersion is observed in the approximate Seeds TNN, where the range of obtained accuracies (i.e., maximum minus minimum accuracy measured) in the Monte Carlo analysis is only $13$\%, while the average accuracy range across all datasets is just $4$\%.
The worst-case accuracy drop compared to the respective ideal (variation-free) TNN is, on average, $7$\%.
Fig.~\ref{fig:variation} demonstrates that even with the very high process variation inherent in printed electronics, the impact on classification accuracy is not catastrophic and remains well bounded in most cases.
Moreover, since in our work the classifier is purely digital, the impact of process variation is significantly smaller compared to the current state-of-the-art analog printed classifiers~\cite{nas_iccad}.

Although variability-induced effects are shown to be well bounded in our TNNs, approaches such as Monte Carlo simulations or variation-aware training may be used to enhance our system's tolerance to process.
Numerous approaches exist that include various robustness-aware training methodologies that account for manufacturing variations~\cite{nas_iccad}, aging~\cite{Zhao:ICCAD2022}, reliability~\cite{Zhao:DATE2023}, and sensing uncertainty~\cite{Zhao:IDP:2023:variability_pe} during the design process~\cite{Tahoori:ETS2025}.
While addressing process variation mitigation is beyond the scope of our work, such methods are orthogonal to our approach and can be seamlessly integrated.}

\subsection{Execution Time Discussion}
Training our arbitrary input precision TNNs took a maximum of $25$ minutes.  
After training, we identify the required LTG and popcount configurations; if any are missing from our library, we generate the corresponding approximate components.  
Fairly large LTGs are required with $n \cdot k$ reaching up to $89$.  
For each LTG approximation (i.e., threshold $\tau$), we perform three independent CGP runs, lasting between $30$min to $2$h, based on the exact LTG's complexity ($k\cdot n$).
Similar timings are used to generate the approximate popcount units.
However, all LTG and popcount approximations run concurrently, not impeding the scalability of our library generation.  
Finally, our multi-objective optimization, which approximates the TNN using our library of approximate components and NSGA-II, is very fast as it performs all evaluations in Python without requiring time-consuming hardware synthesis or simulations.  
In the worst case, it required only $6$ minutes for Pendigits with $4$-bit inputs.  
Experiments ran on a AMD Ryzen 5 3600 with 32GB RAM.  

%% file: tab_soa_cmp.tex
\begin{table}[t!]
\caption{Comparison against the State of the Art.}\label{tab:comp}
\renewcommand{\arraystretch}{1}
\setlength\tabcolsep{4pt}
 \begin{threeparttable}
\begin{tabular}{ll|c|c|c|rr}
\toprule
 \textbf{Dataset} & \textbf{Technique} & \textbf{$\boldsymbol{m}^\ddagger$} & \textbf{BW$^\dagger$} &{\thead{\textbf{Acc$^\mathsection$}\\ (\%)}} &  {\thead{\textbf{Area$^*$} \\ (cm$^{2}$)}}  &  {\thead{\textbf{Power$^*$}\\ (mW)}} \\
\hline 

Arrhythmia & Exact MLP~\cite{Mubarik:MICRO:2020:printedml} & 5 & 4b & 62 & 332 & 1083 \\
Arrhythmia & Exact TNN & 3 & 1b & 60 & 10.2 & 8.39 \\
Arrhythmia & Ax. TNN~\cite{mrazek:iccad:2024} & 3 & 1b & 60 & 9.09 & 7.42 \\
Arrhythmia & \oursTwo{Ours Ax. TNN}$^\star$ & 3 & 1b & 60 & 4.66 & 3.60 \\
Arrhythmia & Ax. MLP~\cite{Afentaki:ICCAD2023:axmac} & 5 & 4b & 60 & 79.9 & 97.7 \\
Arrhythmia & Ax. TNN~\cite{mrazek:iccad:2024} & 3 & 1b & 57 & 6.56 & 5.25 \\
Arrhythmia & \oursFive{Ours Ax. TNN}$^\star$ & 3 & 1b & 57 & 3.85 & 2.81 \\
\hline
BreastCancer & Exact MLP~\cite{Mubarik:MICRO:2020:printedml} & 3 & 4b & 98 & 14.4 & 43.1 \\
BreastCancer & Exact TNN & 6 & 1b & 98 & 0.24 & 0.20 \\
BreastCancer & Ax. TNN~\cite{mrazek:iccad:2024} & 10 & 1b & 98 & 0.10 & 0.05 \\
BreastCancer & \oursTwo{Ours Ax. TNN} & 6 & 1b & 97 & 0.07 & 0.03 \\
BreastCancer & Ax. MLP~\cite{Afentaki:ICCAD2023:axmac} & 3 & 4b & 97 & 2.79 & 3.5 \\
BreastCancer & Ax. TNN~\cite{mrazek:iccad:2024} & 10 & 1b & 93 & 0.09 & 0.05 \\
BreastCancer & \oursFive{Ours Ax. TNN} & 6 & 1b & 95 & 0.06 & 0.02 \\
BreastCancer & Ax. MLP~\cite{Afentaki:DATE2024:gatrain} & 3 & 4b & 94 & 2.45 & 3.13 \\
\hline
Cardio & Exact MLP~\cite{Mubarik:MICRO:2020:printedml} & 3 & 4b & 88 & 38.5 & 131 \\
Cardio & Exact TNN  & 3 & 3b & 89 & 4.22 & 4.87 \\
Cardio & \oursTwo{Ours Ax. TNN} & 3 & 3b & 88 & 2.89 & 3.42 \\
Cardio & Ax. MLP~\cite{Afentaki:DATE2024:gatrain} & 3 & 4b & 87 & 6.55 & 8.21 \\
Cardio & Exact TNN & 3 & 1b & 85 & 0.85 & 0.93 \\
Cardio & Ax. TNN~\cite{mrazek:iccad:2024} & 3 & 1b & 84 & 0.37 & 0.32 \\
Cardio & \oursFive{Ours Ax. TNN} & 3 & 1b & 84 & 0.20 & 0.12 \\
Cardio & Ax. MLP~\cite{Armeniakos:TC2023:codesign} & 3 & 4b & 83 & 10.0 & 23.3 \\
\hline

Pendigits & Exact MLP~\cite{Mubarik:MICRO:2020:printedml} & 5 & 4b & 94 & 70.9 & 218.0 \\
Pendigits & Exact TNN  & 43 & 3b & 94 & 46.8 & 46.8 \\
Pendigits & \oursTwo{Ours Ax. TNN} & 43 & 3b & 92 & 34.2 & 34.2 \\
Pendigits & Ax. MLP~\cite{Armeniakos:TC2023:codesign} & 5 & 4b & 92 & 33.1 & 93.8 \\
Pendigits & Exact TNN  & 41 & 2b & 90 & 31.7 & 32.0 \\
Pendigits & \oursFive{Ours Ax. TNN} & 41 & 2b & 89 & 23.4 & 23.4 \\
Pendigits & Ax. MLP~\cite{Afentaki:ICCAD2023:axmac} & 5 & 4b & 90 & 29.0 & 31.6 \\
\hline
RedWine & Exact MLP~\cite{Mubarik:MICRO:2020:printedml} & 2 & 4b & 56 & 20.3 & 76.9 \\
RedWine & Exact TNN & 3 & 1b & 56 & 0.13 & 0.10 \\
RedWine & Ax. TNN~\cite{mrazek:iccad:2024} & 3 & 1b & 56 & 0.08 & 0.04 \\
RedWine & \oursTwo{Ours Ax. TNN} & 3 & 1b & 57 & 0.06 & 0.02 \\
RedWine & Ax. MLP~\cite{Afentaki:ICCAD2023:axmac} & 2 & 4b & 55 & 2.70 & 3.43 \\
\hline
Seeds & Exact MLP~\cite{Mubarik:MICRO:2020:printedml} & 3 & 4b & 94 & 11.6 & 47.2 \\
Seeds & Exact TNN  & 5 & 2b & 94 & 1.20 & 1.34 \\
Seeds & \oursTwo{Ours Ax. TNN} & 5 & 2b & 92 & 0.49 & 0.42 \\
Seeds & Ax. MLP~\cite{Armeniakos:TC2023:codesign} & 3 & 4b & 92 & 30.9 & 91.0 \\
Seeds & Ax. MLP~\cite{Armeniakos:TC2023:codesign} & 3 & 4b & 89 & 24.1 & 72.9 \\
\hline
WhiteWine & Exact MLP~\cite{Mubarik:MICRO:2020:printedml} & 4 & 4b & 54 & 33.9 & 130 \\
WhiteWine & Exact TNN  & 12 & 3b & 52 & 1.97 & 2.35 \\
WhiteWine & \oursTwo{Ours Ax. TNN} & 12 & 3b & 52 & 1.14 & 1.38 \\
WhiteWine & Ax. MLP~\cite{Armeniakos:TC2023:codesign} & 4 & 4b & 53 & 9.14 & 24.7 \\
WhiteWine & Exact TNN & 11 & 1b & 50 & 0.21 & 0.19 \\
WhiteWine & Ax. TNN~\cite{mrazek:iccad:2024} & 11 & 1b & 50 & 0.16 & 0.13 \\
WhiteWine & \oursFive{Ours Ax. TNN} & 11 & 1b & 50 & 0.06 & 0.02 \\
WhiteWine & Ax. MLP~\cite{Afentaki:ICCAD2023:axmac} & 4 & 4b & 50 & 2.92 & 3.66 \\
\hline
Vertebral & Exact MLP~\cite{Mubarik:MICRO:2020:printedml} & 3 & 4b & 83 & 10.3 & 43.8 \\
Vertebral & Exact TNN  & 27 & 2b & 85 & 3.42 & 3.66 \\
Vertebral & \oursTwo{Ours Ax. TNN} & 27 & 2b & 83 & 0.43 & 0.36 \\
Vertebral & Ax. MLP~\cite{Armeniakos:TC2023:codesign} & 3 & 4b & 81 & 2.85 & 7.46 \\
Vertebral & \oursFive{Ours Ax. TNN} & 27 & 2b & 78 & 0.36 & 0.29 \\

\bottomrule
\end{tabular}
\begin{tablenotes}[flushleft]
      \item[]$^\star$Our TNNs in blue (green) satisfy the $2$\% ($5$\%) accuracy loss constraint. 
      \item[]$^\ddagger$\#Hidden neurons. $^\dagger$Input Precision. $^\mathsection$Test Accuracy.
      \item[]$^*$Area and power, including ADC costs. 
    \end{tablenotes}
 \end{threeparttable}

\end{table}

%% file: conclusion.tex
\section{Conclusions}
Printed electronics, particularly classifier circuits, offer transformative potential in applications requiring flexible substrates and ultra-low costs.
Despite recent advancements, practical implementation faces challenges due to low integration density and limited power from printed batteries and harvesters.
Our research approaches these issues holistically, focusing on optimizations from the sensor-processor interface to the processor itself.
To our knowledge, we present the first open-source end-to-end digital printed classifier that meets both resource and power constraints.
Specifically, we propose an automated framework for designing hardware-efficient approximate printed TNNs by co-optimizing the analog front-end and digital classifier.
Supporting arbitrary input precision and holistic approximation, we show the importance of approximating neurons as a single function and the entire network overall.
Our approach enables fine-tuned area-accuracy trade-offs while minimizing analog-to-digital interfacing overhead, crucial in sensor-based printed applications.
Our TNNs achieve $88$x lower area than the state-of-the-art exact baseline, making them ideal for resource-limited environments, being the only solution enabling printed-battery-powered operation with up to $5$\% accuracy loss.

%% file: bios.tex
\begin{IEEEbiography}[{\includegraphics[width=1in,height=1.25in,clip,keepaspectratio]{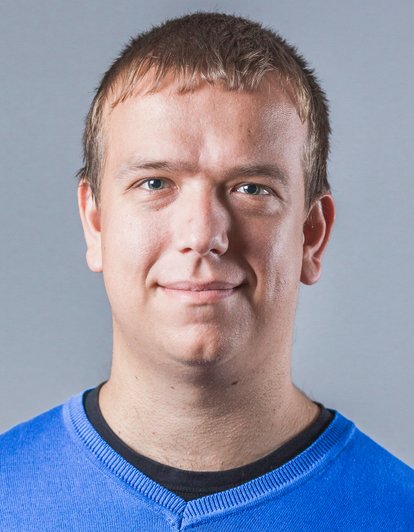}}] {Vojtech Mrazek} is an Assistant Professor at the Brno University of Technology. He received M.Sc. and Ph.D. degrees in information technology from the Faculty of Information Technology, Brno University of Technology, Czech Republic, in 2014 and 2018, respectively. Before he was also a visiting post-doc researcher at Institute of Computer Engineering, Technische Universität Wien (TU Wien), Vienna, Austria (2018-2019). His research interests are approximate computing, genetic programming, formal verification and machine learning. He has authored or co-authored over 60 conference/journal papers focused on approximate computing and evolvable hardware. 
\end{IEEEbiography}

\begin{IEEEbiography}[{\includegraphics[width=1in,height=1.25in,clip,keepaspectratio]{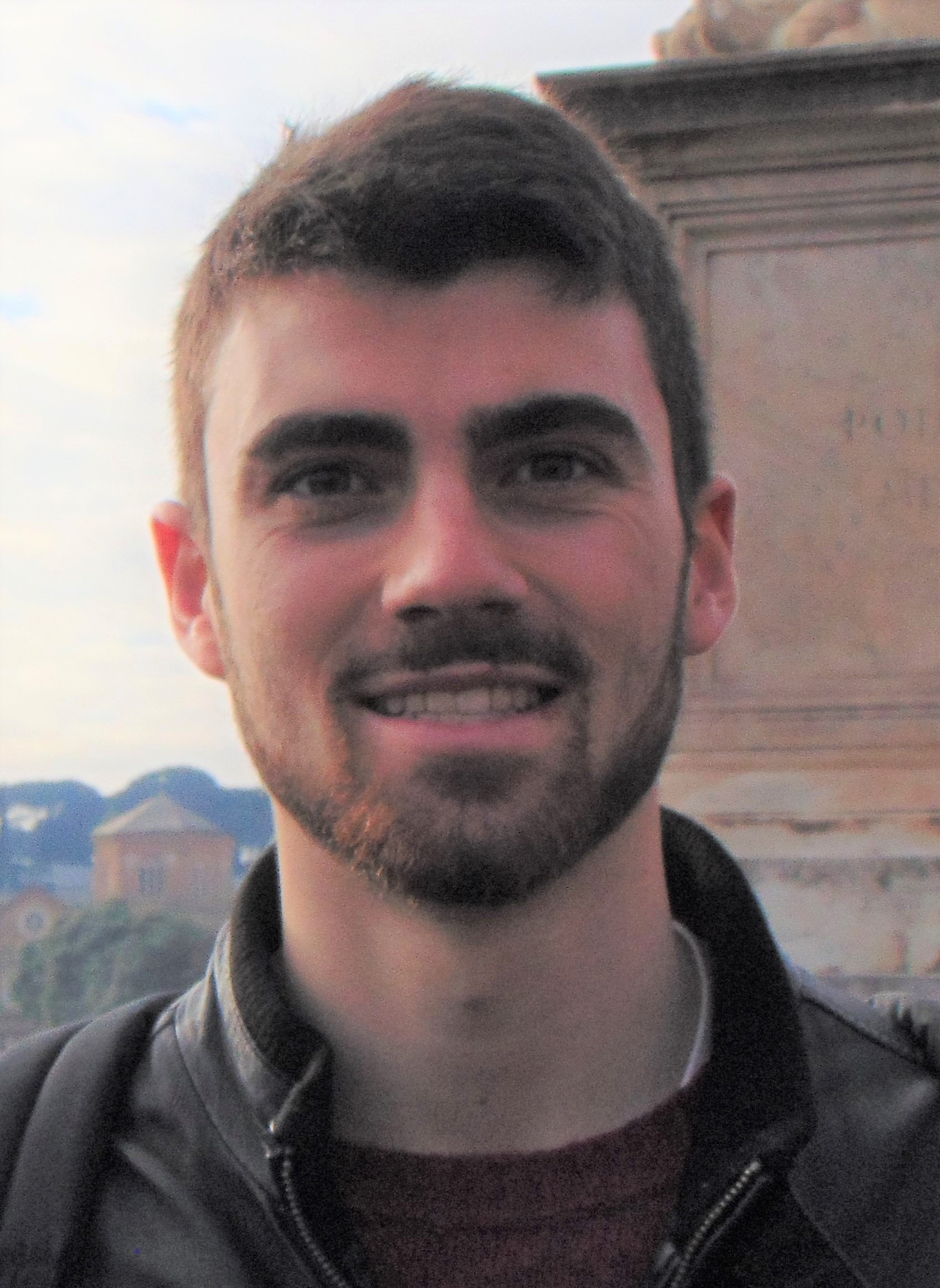}}] {Konstantinos Balaskas} is a post-doctoral researcher at the University of Patras. He received his Ph.D. degree from the Aristotle University of Thessaloniki in 2024, and was a research associate at the Chair for Embedded Systems (CES), at the Karlsruhe Institute of Technology (KIT), between 2021 and 2024. He received his Bachelor Degree in Physics and Master Degree in Electronic Physics from the Aristotle University of Thessaloniki in 2018 and 2020, respectively. His main research interests include embedded machine learning, electronic design automation and physical-driven approximate computing.
\end{IEEEbiography}

\begin{IEEEbiography}[{\includegraphics[width=1in,height=1.25in,clip,keepaspectratio]{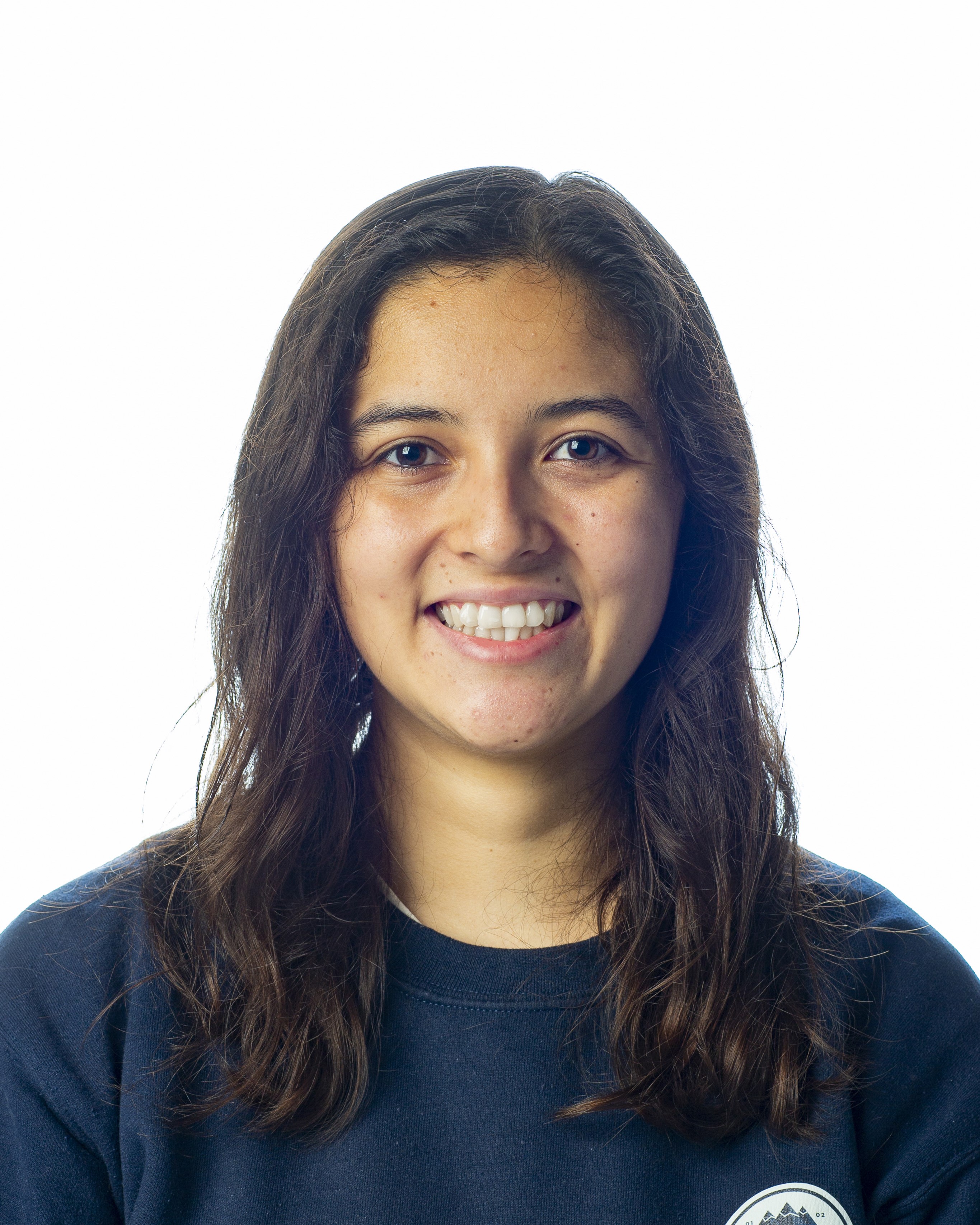}}] {Paula Carolina Lozano Duarte} is a Ph.D. student at the Chair of Dependable Nano-Computing at the Karlsruhe Institute of Technology, Germany. She received her Bachelor's degree (2021) and Master's degree (2023) in Telecommunications Engineering from the Public University of Navarra, Spain. Her main research interests include printed and flexible electronics, machine learning classifiers, and neuromorphic computing.
\end{IEEEbiography}

\begin{IEEEbiography}[{\includegraphics[width=1in,height=1.25in,clip,keepaspectratio]{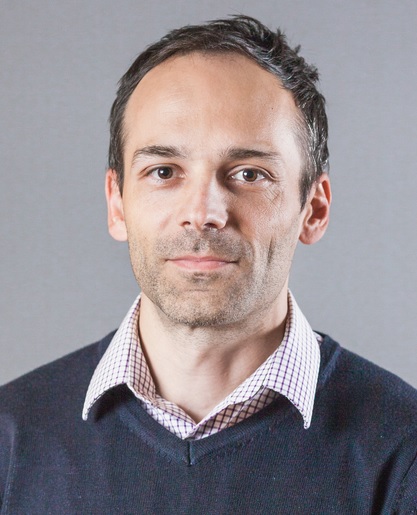}}] {Zdenek Vasicek}
received all his degrees from Brno University of Technology, Czech Republic, where he is currently an Associate professor. He holds a Ph.D. (2012) and an M.S. equivalent (2006) in Computer Science and Engineering. His research interests include formal verification techniques and application of evolutionary approaches in areas related to the design and optimization of complex digital circuits and systems. He is an active PC member of several evolutionary conferences such as EuroGP, GECCO, and ICES. Dr. Vasicek received the Silver and Gold medals at HUMIES, in 2011 and 2015, respectively. 
\end{IEEEbiography}

\begin{IEEEbiography}[{\includegraphics[width=1in,height=1.25in,clip,keepaspectratio]{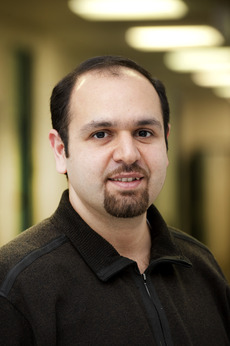}}] {Mehdi B. Tahoori} (M'03, SM'08, F'21) is a Professor at the Chair of Dependable Nano-Computing at Karlsruhe Institute of Technology, Germany. He received the B.S. degree in computer engineering from Sharif University of Technology, Iran, in 2000, and the M.S. and Ph.D. degrees in electrical engineering from Stanford University, Stanford, CA, in 2002 and 2003, respectively. He is currently the deputy editor-in-chief of IEEE Design and Test Magazine. He was the editor-in-chief of Microelectronic Reliability journal. He was the program chair of VLSI Test Symposium in (VTS) 2021 and 2018, and General Chair of European Test Symposium (ETS) in 2019. He is the chair of the IEEE European Test Technology Technical Council (eTTTC). Prof. Tahoori was a recipient of the US National Science Foundation Early Faculty Development (CAREER) Award in 2008. He has received a number of best paper nominations and awards at various conferences and journals. He is a recipient of European Research Council (ERC) Advanced Grant.
\end{IEEEbiography}

\begin{IEEEbiography}[{\includegraphics[width=1in,height=1.25in,clip,keepaspectratio]{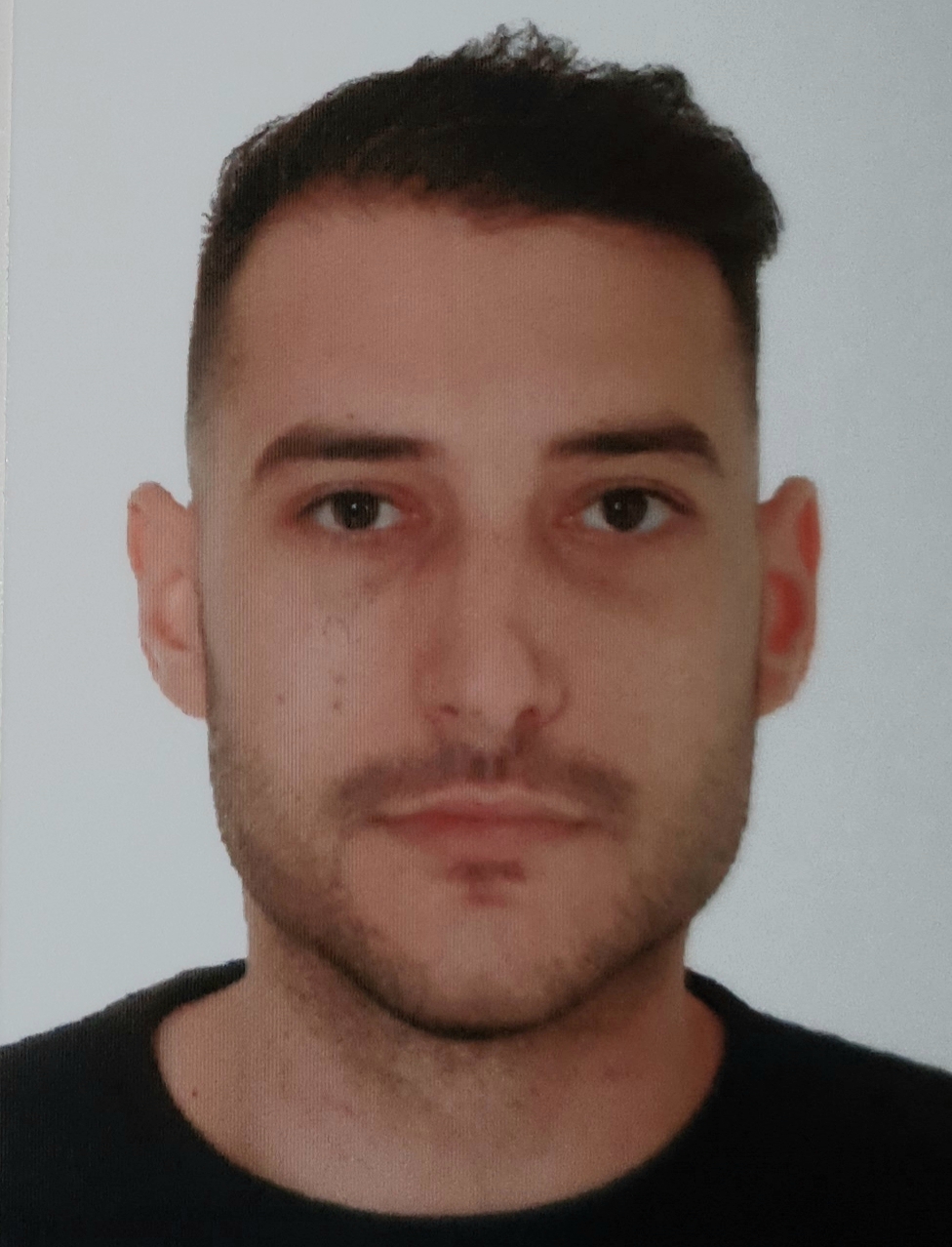}}] {Georgios Zervakis} is an Assistant Professor at the University of Patras. Before that he was a Research Group Leader at the Chair for Embedded Systems (CES), at the Karlsruhe Institute of Technology (KIT) from 2019 to 2022. He received the Diploma and Ph.D. degrees from the School of Electrical and Computer Engineering (ECE), National Technical University of Athens (NTUA), Greece, in 2012 and 2018, respectively. Dr. Zervakis serves as a reviewer in many IEEE and ACM journals and is also a member of the technical program committee of several major design conferences. He has received one best paper nomination at DATE 2022. His main research interests include low-power design, accelerator microarchitectures, approximate computing, and machine learning.
\end{IEEEbiography}